 \def\versionno{adsfamily -- version 1.0 }

\catcode`\@=11

\expandafter\ifx\csname draftcontrol\endcsname\relax\global\def\draftcontrol{0}
\fi

{\count255=\time\divide\count255 by 60
\xdef\hourmin{\number\count255}
\multiply\count255 by-60\advance\count255 by\time
\xdef\hourmin{\hourmin:\ifnum\count255<10 0\fi\the\count255}}
\def\draftdate{\number\month/\number\day/\number\year\ \ \ \hourmin }

\newcommand\makepapertitle{\par
  \begingroup
    \renewcommand\thefootnote{\@fnsymbol\c@footnote}%
    \def\@makefnmark{\rlap{\@textsuperscript{\normalfont\@thefnmark}}}%
    \long\def\@makefntext##1{\parindent 1em\noindent
            \hb@xt@1.8em{%
                \hss\@textsuperscript{\normalfont\@thefnmark}}##1}%
     \newpage
     \global\@topnum\z@   
     \@makepapertitle
     \thispagestyle{empty}\@thanks
  \endgroup
  \setcounter{footnote}{0}%
  \global\let\thanks\relax
  \global\let\makepapertitle\relax
  \global\let\@makepapertitle\relax
  \global\let\@thanks\@empty
  \global\let\@author\@empty
  \global\let\@date\@empty
  \global\let\@title\@empty
  \global\let\title\relax
  \global\let\author\relax
  \global\let\date\relax
  \global\let\and\relax
  \def\version{\let\version\@version\@gobble}
}
\def\@makepapertitle{%
  \newpage
   \ifnum\draftcontrol=1 {}
   \version\versionno
   \vskip 3em%
   \else
   \hfill\hbox to 3cm {\parbox{4cm}{\@pubnum}\hss}%
   \vskip 3em%
   \fi
   \begin{center}%
   \let \footnote \thanks
     {\LARGE \@title \par}%
     \vskip 1.5em%
     {\normalsize
       \lineskip .5em%
       \begin{center} 
         \@author
       \end{center} 
\par}%
     \vskip 1em%
     {\@bstract}%
     \end{center}%
     \vskip .5em
     \@date%
   \par
}

\gdef\@pubnum{}
\def\pubnum#1{%
  \gdef\@pubnum{#1}}

\gdef\@bstract{}
\def\Abstract#1{%
  \gdef\@bstract{%
   \parbox{\textwidth-0pc}{%
   \centerline{\bf Abstract}\penalty1000
   \noindent
   \renewcommand\baselinestretch{1.0}
   {#1}}}
}

\def\ps@paper{\let\@mkboth\@gobbletwo%
     \ifnum\draftcontrol=1
        \def\@oddfoot{\hbox to \textwidth{\tiny \versionno \hfil\tiny\draftdate}%
        \hskip -\textwidth \hbox to \textwidth{\hfil\rm\thepage\hfil}}%
     \else\def\@oddfoot{\hbox to \textwidth{\hfil\rm\thepage\hfil}}
     \fi
     \let\@evenfoot\@oddfoot
}

\def\body{\clearpage
          \pagestyle{paper}
        }
\newenvironment{acknowledgments}{%
\vskip 3.25ex
\noindent {\bf Acknowledgments}
}

\def\@version#1{\ifnum\draftcontrol=1
\typeout{}\typeout{#1}\typeout{}
\vskip3mm\centerline{\hbox{\fbox{\normalsize{\tt DRAFT -- #1 -- }
                   {\draftdate}}}}\vskip3mm
\fi}
\let\version\@version
\long\def\eqlabel#1{\ifnum\draftcontrol=1
                    \tag@false  
                    \tag*{(\theequation) \hbox to -0.2cm{\hspace{0cm}\small{#1}\hss}}
                    \refstepcounter{equation} 
                    \edef\@currentlabel{\theequation}
                    \ltx@label{#1}          
                    \else
                    \label{#1}
                    \fi
                    }
\let\st@bibitem\@bibitem
\let\st@lbibitem\@lbibitem
\ifnum\draftcontrol=1
  \def\@bibitem#1{%
    \st@bibitem{#1}\a@@label{#1}\ignorespaces}
  \def\@lbibitem[#1]#2{%
    \st@lbibitem[#1]{#2}\a@@label{#2}\ignorespaces}
  \def\a@@label#1{%
    \gdef\a@lab{\smash{\normalfont\small#1}}
    \ifvmode
      \if@inlabel
        \global\setbox\@labels\hbox{%
          \llap{\a@lab\let\a@lab\relax
                \kern\@totalleftmargin\kern\marginparsep}%
          \box\@labels}%
      \fi
    \fi}
\fi

\documentclass[12pt,letterpaper]{article}

\usepackage{amsmath,amssymb,array,calc,rotating,epsfig,psfrag}
\usepackage{hyperref}

\ifnum\draftcontrol=1
\fi

\renewcommand\baselinestretch{1.25}
\renewcommand\section{\@startsection {section}{1}{\z@}%
                                   {-3.5ex \@plus -1ex \@minus -.2ex}%
                                   {2.3ex \@plus.2ex}%
                                   {\normalfont\large\bfseries}}
\renewcommand\subsection{\@startsection{subsection}{2}{\z@}%
                                     {-3.25ex\@plus -1ex \@minus -.2ex}%
                                     {1.5ex \@plus .2ex}%
                                     {\normalfont\normalsize\bfseries}}
\renewcommand\subsubsection{\@startsection{subsubsection}{3}{\z@}%
                                     {-3.25ex\@plus -1ex \@minus -.2ex}%
                                     {1.5ex \@plus .2ex}%
                                     {\normalfont\normalsize\it}}

\numberwithin{equation}{section}

\def\projective   {{\mathbb P}}

\def\reals        {{\mathbb R}}
\def\zet          {{\mathbb Z}}

\def\revise#1       {\marginpar{\rule{2mm}{1cm} #1}}

\def\ZZ{\zet}
\def\RR{\reals}
\def\PP{\projective}
\def\RP{\RR\PP}

\def\R{{\rm R}}

\def\sqr#1#2{{\vcenter{\vbox{\hrule height.#2pt  
 \hbox{\vrule width.#2pt height#1pt \kern#1pt
 \vrule width.#2pt}\hrule height.#2pt}}}}

\def\yboxit#1#2{\vbox{\hrule height #1 \hbox{\vrule width #1
\vbox{#2}\vrule width #1 }\hrule height #1 }}
\def\fillbox#1{\hbox to #1{\vbox to #1{\vfil}\hfil}}
\def\ybox{{\lower 1.3pt \yboxit{0.4pt}{\fillbox{8pt}}\hskip-0.2pt}}

\def\comments#1{}

\def\QC{\Bbb{C}}

\def\half{{\frac12}}
\def\pder#1#2{{\frac{\partial{#1}}{\partial{#2}}}}

\def\Re{{\rm Re\hskip0.1em}}
\def\Im{{\rm Im\hskip0.1em}}

\def\ket#1{|#1\rangle}

\def\CN{{\cal N}}

\def\P{\BP}

\def\II{\relax{I\kern-.10em I}}

\def\IZ{\relax\ifmmode\mathchoice
{\hbox{\cmss Z\kern-.4em Z}}{\hbox{\cmss Z\kern-.4em Z}}
{\lower.9pt\hbox{\cmsss Z\kern-.4em Z}}
{\lower1.2pt\hbox{\cmsss Z\kern-.4em Z}}\else{\cmss Z\kern-.4em
Z}\fi}
\def\IB{\relax{\rm I\kern-.18em B}}
\def\IC{{\relax\hbox{$\inbar\kern-.3em{\rm C}$}}}
\def\ID{\relax{\rm I\kern-.18em D}}
\def\IE{\relax{\rm I\kern-.18em E}}
\def\IF{\relax{\rm I\kern-.18em F}}
\def\IG{\relax\hbox{$\inbar\kern-.3em{\rm G}$}}
\def\IGa{\relax\hbox{${\rm I}\kern-.18em\Gamma$}}
\def\IH{\relax{\rm I\kern-.18em H}}
\def\II{\relax{\rm I\kern-.18em I}}
\def\IK{\relax{\rm I\kern-.18em K}}
\def\IP{\relax{\rm I\kern-.18em P}}

\def\Gslash{\relax G\kern-.6em \slash}

\def\inbar{\,\vrule height1.5ex width.4pt depth0pt}

\font\cmss=cmss10 \font\cmsss=cmss10 at 7pt
\def\IR{\relax{\rm I\kern-.18em R}}

\def\BR{\IR}
\def\BP{\IP}
\def\BR{\IR}

\def\Bid{{\mathchoice {\rm {1\mskip-4.5mu l}} {\rm
{1\mskip-4.5mu l}} {\rm {1\mskip-3.8mu l}} {\rm {1\mskip-4.3mu l}}}}

\def\lp10{l_P^{10}}
\def\lp11{l_P^{11}}

\newcommand{\nc}{\newcommand}
\nc{\rnc}{\renewcommand}
\nc{\CY}{Calabi-Yau}
\nc{\CYM}{Calabi-Yau manifold}
\nc{\CYMs}{Calabi-Yau manifolds}
\nc{\DB}{D-Brane}
\nc{\DBs}{D-Branes}
\nc{\SUSY}{supersymmetry}
\nc{\Kah}{K\"ahler}
\nc{\cs}{complex structure}
\nc{\beq}{\begin{equation}}
\nc{\eeq}{\end{equation}}
\nc{\beqa}{\begin{eqnarray}}
\nc{\eeqa}{\end{eqnarray}}
\nc{\ntwo}{${\cal N}=2$}
\nc{\nOne}{${\cal N}=1$}
\nc{\hs}{\hspace{0.2in}}
\nc{\Z}{{\mathbb Z}}
\rnc{\P}{{\mathbb P}}
\rnc{\RP}{{\mathbb {RP}}}
\nc{\WP}{\mathbb{WP}}
\nc{\slag}{special Lagrangian}
\nc{\cn}{\C^n}
\nc{\rn}{\R^n}
\def\ket#1{|#1\rangle}

\nc{\SO}{SO}
\nc{\Sp}{Sp}
\nc{\SU}{SU}

\nc{\Wtree}{W_{\mathrm tree}}
\nc{\Weff}{W_{\mathrm eff}}

\catcode`\@=12

\begin{document}

\title{\Large \bf Holographic  Duals of a Family \break of $\CN=1^*$ Fixed Points}

\pubnum{
hep-th/0506206}
\date{June, 2005}

\author{N. Halmagyi$^\flat$, K. Pilch$^\flat$, C. R\"omelsberger$^\natural$, 
N.P. Warner$^\flat$\\[0.4cm]
\it ${}^\flat$Department of Physics and Astronomy\\
\it University of Southern California \\
\it Los Angeles, CA 90089-0484, USA \\[0.2cm]
\it ${}^\natural$Perimeter Institute \\
\it Waterloo Ontario \\
\it N2L 2Y5, Canada\\[0.2cm]
}

\Abstract{We construct a family of warped $AdS_5$ compactifications
of IIB supergravity that are the holographic duals of the complete set of 
$\CN=1^*$ fixed points  of a $\ZZ_2$ quiver gauge theory. This family 
interpolates between the $T^{1,1}$ compactification with no three-form flux
and the $\ZZ_2$ orbifold of the Pilch-Warner geometry which contains 
three-form flux. This family of solutions is constructed by making the 
most general Ansatz allowed by the symmetries of the field theory.
We use Killing spinor methods because the symmetries
impose two simple projection conditions on the Killing spinors, and these
greatly reduce the problem.  We see that  generic interpolating solution 
has a nontrivial dilaton in the internal  five-manifold. We calculate the central 
charge of the gauge theories from  the supergravity backgrounds and find 
that it is $\frac{27}{32}$ of the parent $\CN=2$, quiver gauge theory. We believe 
that the projection conditions that we derived here will be useful for a much larger 
class of  $\CN=1$ holographic RG-flows.}

\enlargethispage{1.5cm}

\makepapertitle

\vfill \eject 

\tableofcontents

\body

\version\versionno

\section{Introduction}

Motivated by the AdS/CFT duality 
\cite{Maldacena:1997re,Witten:1998qj}, there has been considerable
interest in finding explicit supergravity solutions that correspond to
conformal field theories with N=1 supersymmetry in four dimensions. In
this paper we solve a long standing problem in this context which was
originally posed in \cite{Corrado:2002wx}:  We find the conjectured
family of solutions that correspond to infra-red fixed-points that interpolate
between the solution of Pilch and Warner \cite{Pilch:2000ej}, and the solution
of Romans  \cite{Romans:1984an} that is the basis of what has become
known as the Klebanov-Witten (KW) point  \cite{Klebanov:1998hh}. 

To be more precise, the Klebanov and Witten \cite{Klebanov:1998hh}  argued 
that if one starts with the $\CN=2$, four-dimensional, $\widehat{A}_1$ quiver 
gauge theory  and breaks it to an $\CN=1$ supersymmetric field theory by 
introducing a (unique) $SO(4)$ invariant superpotential, the theory will flow 
to an  $\CN=1$ superconformal fixed point in the infra-red and this fixed point
is dual to the solution of IIB supergravity on ${\rm AdS}_5\times T^{1,1}$ 
\cite{Romans:1984an}.   Similarly, it was argued in \cite{Freedman:1999gp, 
Corrado:2002wx} that the same $\widehat{A}_1$ quiver gauge theory would, 
under a  particular $SO(3)$ invariant  superpotential,  flow to another $\CN=1$ 
superconformal fixed point whose supergravity dual  is  the ($\ZZ_2$ orbifold of)
the Pilch-Warner (PW) solution \cite{Pilch:2000ej} whose existence was first 
discovered via five-dimensional  supergravity \cite{Khavaev:1998fb}.  

More generally, it was argued in  \cite{Corrado:2002wx}, using the non-perturbative 
methods of Leigh and Strassler  \cite{Leigh:1995ep}, that there is a family of
four-dimensional $\CN=1$ superconformal field theories (SCFT's) that continuously
interpolate between the KW  flow and the PW flow, and that this family preserves at 
least an $SO(3)$ global symmetry.    Indeed, \cite{Corrado:2002wx} also investigated 
the corresponding five-dimensional gauged supergravity 
solutions that were expected to capture the relevant sectors of the IIB 
supergravity dual of the family of flows.  From the five-dimensional perspective, the 
existence of the family of flows and of the family of IR fixed points was almost a triviality.
There was, however, an important caveat:  There are no consistent truncation
theorems for this more general class of five-dimensional supergravity theories,
and so the five-dimensional result were very suggestive, but did not prove that
there had to be corresponding ten-dimensional solutions.   The search for this 
family of solutions within IIB supergravity  has been rather long and surprisingly 
difficult, and here we will prove that family exists by reducing the problem to a 
system of ordinary differential equations and exhibiting numerical solutions.

Much of the technology for finding supersymmetric solutions to supergravities in 
various dimensions relies on solving the supersymmetry variations and the Bianchi 
identities, one can {\it a postieri} check that the field equations are satisfied.  
A general formalism for analyzing the Killing spinor equations is that of 
$G$-structures.    For IIB supergravity, this works extremely well when the
internal manifold has $SU(3)$ structure \cite{Gauntlett:2004zh,Butti:2004pk} but for backgrounds with 
only $SU(2)$ structure (which is the structure appearing in the current work) that 
methodology is too cumbersome at present \cite{Dall'Agata:2004dk}.   A more pragmatic 
approach developed by two of the current authors and their collaborators 
\cite{Gowdigere:2003jf, Pilch:2003jg, Pilch:2004yg}, is to use the physics of the
problem to make an Ansatz for the  Killing spinors as well as the metric and form 
fields.   We will follow the latter approach  and find that the symmetries of the 
problem sufficiently restrict the form of the Ansatz  such that the full solution can be 
obtained.  More specifically, the``supersymmetry bundle'' is a four-dimensional  
subspace of the of the 32 real components of the spinors, and we can use the 
symmetries and a specific combination of the gravitino variation equations to 
define an eight-dimensional subspace that contains the Killing 
spinors.   We then parametrize the supersymmetries within this eight-dimensional 
subspace in a manner  that is equivalent to the dielectric deformation of the 
canonical $D3$-brane  projector \cite{Gowdigere:2003jf, Pilch:2003jg, Pilch:2004yg}.
Having found the supersymmetries, one can then build the rest of the solution from the Killing 
spinor equation.

The  solution of IIB supergravity on ${\rm AdS}_5\times T^{1,1}$  is a 
Freund-Rubin Ansatz with constant dilaton-axion and vanishing three-form flux. One
can re-cast this solution in terms of $D3$ branes  on the conifold, and the metric
transverse  to the branes is thus K\"ahler and Ricci flat.  It therefore possesses 
a rather trivial $SU(3)$ structure.  The PW solution is a warped Freund-Rubin Ansatz 
with constant dilaton-axion and non-vanishing three-form flux. The PW metric is 
neither Ricci flat nor K\"ahler but it is equipped with an integrable complex structure, 
namely that of $A_{1}\times \QC$ \cite{Halmagyi:2004jy}. It has two globally-defined 
spinors and as such has only $SU(2)$ structure.   We find that the interpolating 
solution also has only an $SU(2)$ structure. The surprise is that even though
 the two end points of our interpolation have a trivial dilaton-axion, the interpolating 
 solutions themselves have a non-trivial dilaton-axion.   It also seems that the
 interpolating family lacks a integrable complex structure.
 
It is worth mentioning the interesting recent work \cite{Lunin:2005jy} in which 
the authors use the eight-dimensional duality group to generate new solutions that can 
be easily lifted to ten dimensions. For the supergravity duals to SCFT's they are able to 
identify the exactly marginal operator in the field theory,  thus providing a holographic 
check of the methods of Leigh and Strassler. Our scenario falls out of the scope of the 
powerful methods employed there since it lacks the required two $U(1)$ 
non-$R$ symmetries.

This paper is organized as follows: In section 2, we review the relevant field theory, 
and in particular discuss the symmetries.  The symmetries of the supergravity background 
are discussed in Section 3. In Section 4, we reduce the problem to five dimensions, 
enforcing the ${\rm AdS}_5$ factor in the ten-dimensional background.  Section 5 contains 
a review of the KW and PW solutions. Sections 6 and 7 contain the main calculations:
We derive the BPS equations from the most general Ansatz which preserves the 
relevant symmetries and reduce this system to three first order, non-linear ordinary differential 
equations.   In section 7 we establish that there is indeed a one parameter family of
regular solutions to these BPS equations and solve them numerically.  We indeed
show that they interpolate between the KW and PW solutions.  Those who are interested
in the main result should therefore jump to sections 6 and 7.   Section 8 contains a 
discussion of the central charge of each gauge theory in the family from the perspective
of the dual supergravity theory. We show analytically that the central charge has the
correct constant value across the entire family of solutions.  Finally, there are several
appendices containing spinor conventions and computational details.

\section{Field theory considerations}

The conformal field theory we are considering in this paper is a
non-trivial IR fixed point of a mass deformed $\CN=2$ quiver
gauge theory \cite{Douglas:1996sw}. The UV field theory has an 
$SU(N)\times SU(N)$ gauge group two
bi-fundamental hypermultiplets, one in the $(N,\bar N)$ and one in 
the $(\bar N,N)$.  In $\CN=1$ language the first hypermultiplet 
decomposes into two chiral multiplets $(A_1,B_1)$ and the second 
hypermultiplet decomposes into two chiral multiplets $(A_2,B_2)$.

The superpotential of this theory is
\beq
W~=~ {\rm Tr} \left (\phi_1(A_1B_1-B_2A_2) \right)~+~ {\rm Tr} \left( 
\phi_2(A_2B_2-B_1A_1) \right).
\eeq
This theory has an $SU(2)\times SU(2)_R\times U(1)_R$ continous 
global symmetry. The two hypermultiplets form a doublet under the 
$SU(2)$ flavor symmetry.

This theory can be deformed by mass terms for the adjoint scalars 
\cite{Leigh:1995ep,Corrado:2002wx,Corrado:2004bz}
\beq
\Delta W ~=~ \frac{m_1}{2} \, {\rm Tr} \left (\phi_1^2\right) ~+~ 
\frac{m_2}{2}\,  {\rm Tr} \left (\phi_2^2\right).
\eeq
This deformation breaks the continuous global symmetry to $SU(2)\times 
U(1)_R$.   The $U(1)_R$ symmetry is actually a combination of the
$U(1)_R$ symmetry and a $U(1)$ subgroup of the $SU(2)_R$ symmetry 
of the $\CN=2$ theory. The R-charges of the fields are
\beq
\begin{array}{c|c|c}
\phi_i & A_i & B_i \\ \hline
1 & \half & \half
\end{array}
\eeq
This field theory also has a $\ZZ_4$ discrete symmetry, with a generator
$\omega$ which acts as a charge conjugation
\beqa
\phi_i&\mapsto&\phi_i^t \\
A_i&\mapsto&iA_{i+1}^t \\
B_i&\mapsto&iB_{i+1}^t
\eeqa
It is easy to see that the $\ZZ_4$ symmetry commutes with the continous
symmetries and so the global symmetry of the theory is $SU(2)\times\ZZ_4\times 
U(1)_R$.  However, $\omega^2$ and the center of $SU(2)$ simply negate
$A_i$ and $B_i$, and in the supergravity dual we will consider only 
gauge-invariant bilinears of the fields $A,B$.  Thus  these generators will
act trivially in supergravity which means that the symmetry of the supergravity
theory\footnote{Indeed, even within the $SU(N)$ gauge
theory, for $N$ even,  negating $A_i$ and $B_i$ is in the center of the 
$SU(N)$ gauge groups and so the symmetry of (perturbative) physical states of
the field theory  will also be $SO(3)\times\ZZ_2\times U(1)_R$.}  will be  
$SO(3)\times\ZZ_2\times U(1)_R$.

Below the mass scale given by $m_1$ and $m_2$ one can integrate out 
the adjoint scalars $\phi_1$ and $\phi_2$ and the low energy superpotential 
is given by
\beq
W~=~ \lambda_1\,  {\rm Tr} \left ((A_1B_1-B_2A_2)^2\right) ~+~  \lambda_2 \, 
{\rm Tr} \left( (A_2B_2-B_1A_1)^2\right).
\eeq
The low energy effective action has the two gauge couplings $\tau_i$ 
and the two quartic superpotential couplings $\lambda_i$. 

The deformed theory is believed to flow to a non-trivial IR fixed point.
Vanishing of the $\beta$-functions for all the couplings requires
\beq
\gamma_{A_i}(\tau_1,\tau_2,\lambda_1,\lambda_2) ~+~ 
\gamma_{B_i}(\tau_1,\tau_2,\lambda_1,\lambda_2)~+~ \half ~=~ 0.
\eeq
This is two equations for four unknowns. However, the $SU(2)\times\ZZ_4$ 
symmetry implies that the functional form of all the anomalous dimensions 
is the same
\beq
\gamma ~=~ \gamma_{A_i}  ~=~ \gamma_{A_{i+1}} ~=~ \gamma_{B_i} ~=~
\gamma_{B_{i+1}}.
\eeq
From this we conclude that the vanishing of the $\beta$-function implies
only one constraint 
\beq
\gamma ~+~ \frac{1}{4} ~=~ 0
\eeq
for four unknowns.  We expect the moduli space of IR theories to have
three complex dimensions.

The central charges for such theories have been calculated in 
\cite{Gubser:1998vd, Anselmi:1997am, Freedman:1999gp}. 
The ratio of the central charges of the  IR theory and the UV theory is
\beq
\frac{c_{(IR)}}{c_{(UV)}} ~=~ \frac{27}{32}.
\eeq

\section{Realizing the global symmetries within supergravity}

In the following we want to construct supergravity backgrounds that 
are holographic duals to field theories with a given global symmetry 
algebra. The global symmetries have to be realized as symmetries of 
the background and this leads to powerful constraints on the background. 
One set of constraints comes from the 
existence of the symmetry generators. The other set of constraints comes
from the commutation relations of the symmetry generators.

Type IIB supergravity has six different gauge symmetries. General coordinate 
transformations which we restrict to the isometries generated by Killing vectors,
$\delta(\xi)$;  local  Lorentz transformations, $\delta(l)$;  the U(1) R-symmetry,
$\delta(\Sigma)$;  the gauge transformations of the two-form and four-form potentials,
$\delta(\Lambda^{(1)})$ and $\delta(\Lambda^{(3)})$, and the supersymmetry 
transformation $\delta(\epsilon)$. There is also a global $SU(1,1)$ symmetry,
which in string theory is broken to $SL(2,\ZZ)$.

A background has a global symmetry generated by some specific symmetry
generators, $(\xi,l,\Sigma,\Lambda^{(1)},\Lambda^{(3)},\epsilon)$, provided that
this transformation leaves the background invariant\footnote{There are also the
$SL(2,\ZZ)$ actions, but those are discrete symmetries.}.  Global 
supersymmetries have to be generated just by an $\epsilon$ and 
global bosonic symmetries will be generated by a combination 
$(\xi,l,\Sigma,\Lambda^{(1)},\Lambda^{(3)})$.

\subsection{Continous bosonic symmetries}

The non-trivial, physical bosonic symmetries of the background must
involve a  transformation  by an isometry, or a Killing
vector\footnote{One can see this from the fact, that a transformation 
generated by $(\xi=0,l,\Sigma,\Lambda^{(1)},\Lambda^{(3)})$ will not
leave any field configuration invariant.}, $\xi$.
The vielbein only transforms under both general coordinate and local Lorentz 
transformations and its invariance typically requires a compensating
local Lorentz transformation that depends on the choice of the vielbein.
For this reason it is often useful, wherever possible, to choose a vielbein made of 
invariant  one-forms.

The coset fields $V^\alpha_\pm$, which describe the dilaton
and axion, transform under the global $SU(1,1)$ symmetry and locally
under general coordinate transformations and the 
$U(1)$ R-symmetry of the IIB theory.  The invariance of the coset fields requires
\beq\label{invofv}
(\partial_\xi\pm i\Sigma)V^\alpha_\pm=0.
\eeq
From this it is easy to derive that the gradient of the dilaton-axion field
in the $\xi$ direction is vanishing
\beq
P_\xi=0
\eeq
and that $\Sigma$ is the $\xi$ component $Q_\xi$ of the $U(1)$ connection. 
Equation (\ref{invofv}) also implies, that at least one of the $V^\alpha_\pm$ 
is a non-vanishing section of the associated line bundle over an orbit 
of $\xi$. This implies, that  one can choose a trivialization of the
$U(1)$ bundle over an orbit of $\xi$ and thereby render the connection, 
$Q_\xi$,   trivial: 
\beq
Q_\xi=0.
\eeq

With this choice of trivialization the action of the symmetry on the 
field strengths $G^{(3)}$ and $F^{(5)}$ is just the Lie derivative. 
For this reason $G^{(3)}$ and $F^{(5)}$ have to be invariant forms
\beq
\pounds_\xi G^{(3)}=0\qquad{\rm and}\qquad\pounds_\xi F^{(5)}=0.
\eeq

We will not discuss the gauge transformations 
$\delta(\Lambda^{(1)},\Lambda^{(3)})$ here because the supersymmetry 
variations, the Bianchi identities and the equations of motion depend only
on the field strengths $G^{(3)}$ and $F^{(5)}$.

From the above discussion it follows, that the Killing vectors have to
satisfy the bosonic Lie algebra of the global symmetry group of the 
background
\beq
[\delta(u_1),\delta(u_2)]=\delta([u_1,u_2]).
\eeq
This implies that the background is a fibration of a product 
of coset spaces and group manifolds over a possibly non-trivial base.

\subsection{Supersymmetries}

The supersymmetries are generated by Killing spinors $\epsilon$. In a purely
bosonic background the requirement of the existence of a global 
supersymmetry is the vanishing of the dilatino and the gravitino variation.

Before looking at the dilatino and gravitino variation it is 
useful to look at the commutators of the supersymmetry generators with
other symmetry generators.
\beq
[\delta(g),\delta(\epsilon)]=
[\delta(\xi),\delta(\epsilon)]+[\delta(l),\delta(\epsilon)]+
[\delta(\Sigma),\delta(\epsilon)]=
\delta\Big(\Big(\partial_\xi+\frac{1}{4}l_{rs}\gamma^{rs}-\frac{i}{2}Q_\xi\Big)
\epsilon\Big),
\eeq
where $l$ is a ``Lie connection.''  If the vielbein is given in terms of invariant
forms and the U(1) connection is chosen trivially, then the above expression 
reduces to the ordinary derivative. For later convenience we define the Lie 
derivative of $\epsilon$ by this derivative operator:
\beq
\pounds_\xi\epsilon=
\Big(\partial_\xi+\frac{1}{4}l_{rs}\gamma^{rs}-\frac{i}{2}Q_\xi\Big)\epsilon.
\eeq
This gives rise to the differential equation
\beq
\pounds_\xi\epsilon=g\cdot\epsilon
\eeq
This allows to determine the dependence of $\epsilon$ on the directions 
given by the symmetries.

There are also powerful constraints coming from the anti-commutator of
two supersymmetries
\beq
\{\delta(\epsilon_1),\delta(\epsilon_2)\}=\delta(\{\epsilon_1,\epsilon_2\}).
\eeq
This implies, that
\beq
\xi^\mu=2\Im(\bar\epsilon_1\gamma^\mu\epsilon_2)
\eeq
is the Killing vector associated to $\{\epsilon_1,\epsilon_2\}$ and that
\beq
l^{rs}=\omega_\xi{}^{rs}-
\frac{1}{3}F^{rsmnp}\Re(\bar\epsilon_1\gamma_{mnp}\epsilon_2)+
\frac{3}{4}\Im\left(G^{rsm}\bar\epsilon_1\gamma_m\epsilon_2^*+
\frac{1}{18}G_{mnp}\bar\epsilon_1\gamma^{rsmnp}\epsilon_2^*\right)
\eeq
is the local Lorentz transformation associated to $\{\epsilon_1,\epsilon_2\}$.
The relation including $\Sigma$ is trivially satisfied.

\subsection{Discrete symmetries}

Discrete symmetries can be composed out of global diffeomorphisms, 
local Lorentz transformations, gauge transformations for the form fields,
$U(1)_R$ symmetry transformations and $SL(2,\ZZ)$ transformations. The
commutation relations with the other global symmetries are again given
by the field theory.

The constraints from discrete symmetries are especially powerful
when the global diffeomorphism leaves the orbits of the continous 
symmetries invariant. If this is the case, the discrete symmetry
implies powerful projection conditions on the fields and supersymmetry 
generators.  We will see an explicit example of this below.

\section{Reduction to a five-dimensional problem}

\subsection{Decomposing the metric and spinors}

The bosonic part of the four-dimensional, $\CN=1$  superconformal algebra is
$SO(2,4)\times U(1)$. This bosonic symmetry is realized 
by Killing vectors in the ten-dimensional geometry. The geometry is then 
$AdS_5$, which is covering space of $SO(2,4)/SO(1,4)$,  warped over an 
internal five-manifold. The internal five-manifold itself is a $S^1$ 
fibration over a four-manifold, $X_4$.  The $S^1$ must be a Killing direction
dual to the $R$-symmetry action.

We adopt  the following index conventions:  Capital Latin 
letters denote ten-dimensional indices ($0,\cdots,9$), small Greek letters
denote the five-dimesional indices in the $AdS^5$ ($0,\cdots,4$) and small 
Latin letters denote the internal indices ($5,\cdots,9$). A hat denotes 
ten-dimensional frame indices, a tilde denotes five-dimensional frame indices 
in $AdS_5$ and a check denotes five-dimensional frame indices in the internal 
space.  
The warped $AdS_5$ leads to a vielbein Ansatz of the form:
\beqa
e^{\hat\mu}&=&\Omega\,  e_{(e)}^{\tilde\mu} \quad{\rm for}\quad \mu=0,\cdots,4 \\
e^{\hat m}&=&\frac{1}{\Omega\, }e_{(i)}^{\check m} \quad{\rm for}\quad m=5,\cdots,9
\eeqa
where $e_{(e)}^{\tilde\mu}$ is a vielbein for $AdS_5$ of unit curvature 
radius and $e_{(i)}^{\check m}$ is a vielbein for the internal manifold.

The spinors of IIB supergravity must similarly decompose into 
spinors on $AdS_5$ and  on the internal five-manifold.   We will analyze
this in detail, and  we need to recall some basic facts about spinors 
in various dimensions.  More information  may be found in Appendix A.

Recall that in the IIB theory one can impose a Majorana-Weyl
condition on a spinor to reduce it to 16 real components.  It is most
convenient to represent the 32 components of the $\CN=2$ supersymmetry 
of the IIB theory in terms of a complex Weyl spinor.  Our task will
be to decompose this into components along the two five-manifolds. 
To do this it will be important to recall how complex conjugation acts on
spinors.  Given a set of $\gamma$-matrices, complex conjugation maps
them into an equivalent set, and so there is a matrix, $B$, that will
generically conjugate the $\gamma_A^*$ back to the  $\gamma_A$.
By the same token, to map a spinor, $\Psi$, to its
complex conjugate representation one must accompany the conjugation
by the action of $B$.  Thus, the conjugate spinor, $\Psi^\odot$, is defined
by:
\beq\label{conjspinor}
\Psi^\odot ~\equiv~ B^{-1} \Psi^*\,.
\eeq
The form and properties of $B$  depend upon the dimension and
signature of the metric and upon   $\gamma$-matrix conventions. 
In the IIB theory one can adopt conventions in which $B$ is the
identity matrix (as in \cite{Schwarz:1983qr})), but we will keep our expressions
convention independent and adopt the notation (\ref{conjspinor}).
In five Lorentzian dimensions, $B$ is necessarily non-trivial and
may be thought of as a symplectic form.  Indeed, this fact lies at the heart 
of the symplectic Majorana condition of five-dimensional supersymmetric
theories.  

In the following we will use the the notation (\ref{conjspinor}) to denote
the conjugate spinor in all dimensions and metric signatures.

\subsubsection{Killing spinors on $AdS_5$}

The ten-dimensional Killing spinor is a complex, chiral spinor 
($\gamma_{(10)}\epsilon=\epsilon$). Since it has to respect the symmetries
of $AdS_5$, it has to be built out of five-dimensional Killing spinors.
The five-dimensional Killing spinor equation is:
\beq\label{Kspeqn}
(D_\mu\pm\frac{i}{2}\gamma_\mu)\, \zeta=0\,,
\eeq
for either choice of sign.  The distinct signs determine the
transformation properties under the conformal group, $SO(2,4)$.
That is, solutions with a plus (respectively, minus) sign
transform in the $4$ (respectively, $\bar 4$) of $SO(2,4)$.
If $\zeta$ is any $SO(1,4)$ four-spinor satisfying (\ref{Kspeqn}) for one
choice of sign, it is easy to see  that $\zeta^\odot$  is
a solution to (\ref{Kspeqn})   with the opposite sign. 

One can also check that $2\Re(\bar\zeta_1\gamma^\mu\zeta_2)$ are 
Killing vector fields generating the $AdS_5$ group and that 
$\Re(\bar\zeta_1\zeta_2)$ generates the $U(1)_R$ symmetry in 
accordance with the four-dimensional, $\CN=1$  superconformal algebra. 
This is because the superconformal algebra implies that the bosonic symmetry 
generators appear in the $4\otimes\bar 4=1\oplus 15$.
On the other hand, expressions like $\Re(\bar\zeta_1\zeta_2^\odot)$ and 
$\Re(\tilde\zeta_1\gamma^\mu \zeta_2)$ are not related Killing
vectors or other bosonic symmetry generators.  
 
\subsubsection{The ten-dimensional Killing spinors}

The ten-dimensional Killing spinors can be decomposed as
\beq\label{killingspa}
\epsilon_\zeta=\Omega^\half\left(\begin{array}{c} 
\zeta\otimes\chi^{(1)}+
(\zeta^\odot)\otimes(\chi^{(2)}{}^\odot) \\ 0 
\end{array}\right),
\eeq
where $\zeta$ is a Killing spinor in $AdS_5$
which does not depend on the internal coordinates 
and $\chi^{(i)}$ are independent internal five-dimensional spinors which only depend
on the internal coordinates.

We can now compute the Killing vectors  
$\Re(\bar\epsilon_1\gamma^M\epsilon_2)$
\beqa
\Re(\bar\epsilon_1\gamma^\mu\epsilon_2)&=&\Omega\,
\Re(\bar\zeta_1\gamma^\mu\zeta_2)\,
(\bar\chi^{(1)}\chi^{(1)}+\bar\chi^{(2)}\chi^{(2)}),\\
\label{KVpara} \Re(\bar\epsilon_1\gamma^m\epsilon_2)&=&\Omega\,
\Re(\bar\zeta_1\zeta_2)\,
(\bar\chi^{(1)}\gamma^m\chi^{(1)}+\bar\chi^{(2)}\gamma^m\chi^{(2)}).
\eeqa
It is interesting to note that cross terms like 
$\tilde\chi^{(1)}\chi^{(2)}$ cancel out in this expression. The
foregoing equations also give rise to normalization conditions for
the $\chi^{(i)}$. The condition coming from the normalization of 
the Killing vectors parallel to the $AdS_5$  is:
\beq\label{normc1}
\bar\chi^{(1)}\chi^{(1)}+\bar\chi^{(2)}\chi^{(2)}=1.
\eeq

Similarly, the Killing vector of the form (\ref{KVpara}) along the 
internal manifold must be that of the $U(1)_R$ symmetry, and so
we must have:
\beq\label{normc2}
\frac{3}{2}\pder{}{\phi}=\Omega\, (\bar\chi^{(1)}\gamma^m\chi^{(1)}+ 
\bar\chi^{(2)}\gamma^m\chi^{(2)})\, e_{\hat m},
\eeq
where $\phi$ is an internal coordinate. 
 
Finally, we can determine the $\phi$ dependence of the internal spinors
$\chi^{(i)}$. Since the $\phi$ direction realizes the $U(1)_R$ symmetry,
we have to impose
\beq
\pounds_{\pder{}{\phi}}\epsilon_\zeta=\epsilon_{i\zeta},
\eeq
which is equivalent to the five-dimensional spinor $\zeta$ having charge
1 under the $U(1)_R$ symmetry. This leads to
\beq
\pounds_{\pder{}{\phi}}\chi^{(i)}=i\chi^{(i)}.
\eeq

\subsection{The dilatino variation}

The dilatino variation is given by \cite{Schwarz:1983qr}
\beq
\delta\lambda=iP_{\hat M}\gamma^{M}\epsilon^\odot-
\frac{i}{24}G_{\hat M \hat N \hat P}\gamma^{MNP}\epsilon.
\eeq
Poincar\'e invariance requires that $P$ and $G$ only have components
in the  internal directions. This leads to the equation:
\beq
0=P_{\hat m}\gamma^{m}\epsilon^\odot-
\frac{1}{24}G_{\hat m \hat n \hat p}\gamma^{mnp}\epsilon.
\eeq
Inserting the form of the Killing spinor (\ref{killingspa}) and realizing 
that $\zeta$ and $\zeta^\odot$ may be considered as independent
variables, we get the two five-dimensional equations:
\beqa\label{dilvara}
\delta\lambda^{(1)}=P_{\check m}\gamma_{(i)}^{m}\chi^{(2)}
+\frac{\Omega^2}{24}G_{\check m\check n\check p}
\gamma_{(i)}^{mnp}\chi^{(1)}=0,\\
\label{dilvarb}
\delta\lambda^{(2)}=P_{\check m}\gamma_{(i)}^{m}\chi^{(1)}{}^\odot
+\frac{\Omega^2}{24}G_{\check m\check n\check p}
\gamma_{(i)}^{mnp}\chi^{(2)}{}^\odot=0.
\eeqa
Since the background fields are independent of the
$U(1)_R$ direction, these  equations  reduce to  
spinor equations on the four-dimensional  base, $X_4$, of the $S^1$
fibration that makes up the internal manifold.
It is also  easy to show that the component of  $P_{\check m}$ 
along the $U(1)_R$ Killing vector must vanish. This result is expected 
from the $U(1)_R$ invariance, but can be deduced explicitly from
(\ref{dilvara}) and (\ref{dilvarb}) as follows:  Multiply the first equation  by 
$\bar\chi^{(2)}$,  transpose the second  equation and multiply it by $\chi^{(1)}$ 
and add the  two.

\subsection{The gravitino variation}

The gravitino variation is \cite{Schwarz:1983qr}:
\beq
\delta\psi_{\hat M}=D_{\hat M}\epsilon+
\frac{i}{480}F_{\hat P\hat Q\hat R\hat S\hat T}
\gamma^{PQRST}\gamma_{M}\epsilon+
\frac{1}{96}G_{\hat P\hat Q\hat R}\left(\gamma_{M}{}^{PQR}-
9\delta^{P}_{M}\gamma^{QR}\right)\epsilon^\odot,
\eeq
where the covariant derivative is given by
\beq
D_{\hat M}\epsilon=\partial_{\hat M}\epsilon+
\frac{1}{4}\omega_{\hat M\hat P\hat Q}\gamma^{PQ}\epsilon-
\frac{i}{2}Q_{\hat M}\epsilon.
\eeq

In order to continue, we need to determine the ten-dimensional spin connection
in terms of the warp factor $\Omega$ and the five-dimensional spin connection:
\beqa
\omega_{\hat\mu\hat\nu\hat\rho}&=&
\Omega^{-1}\omega_{(e)}{}_{\tilde\mu\tilde\nu\tilde\rho},\\
\omega_{\hat\mu\hat\nu\hat r}&=&\partial_{\check r}\Omega
\eta_{\hat\mu\hat\nu},\\
\omega_{\hat\mu\hat n\hat r}&=&0,\\
\omega_{\hat m\hat\nu\hat\rho}&=&0,\\
\omega_{\hat m\hat\nu\hat r}&=&0,\\
\omega_{\hat m\hat n\hat r}&=&-\Omega\omega_{(i)}{}_{\check m\check n\check r}-
\partial_{\check n}\Omega\delta_{\check m\check r}+
\partial_{\check r}\Omega\delta_{\check m\check n},
\eeqa
where $\omega_{(e)}$ is the spin connection on $AdS_5$ and $\omega_{(i)}$ 
is the spin connection on the internal manifold.   Also note that
\beq
Q_{\hat\mu}=0.
\eeq

The self dual five-form flux can be written as
\beq
F^{(5)}=f\,e^{\hat 0\cdots\hat 4}+f\,e^{\hat 5\cdots\hat 9},
\eeq
where $f$ only depends on the internal coordinates. The Bianchi identity
for $F^{(5)}$ reduces, for such a compactification, to
\beq
dF^{(5)}=0,
\eeq
which implies
\beq
f=\frac{f_0}{\Omega^5},
\eeq
where $f_0$ is an integration constant.

Now we can determine the gravitino variations with $M=0,\cdots,4$ 
a similar argument as for the dilatino variation leads to
\beqa\label{gravvara}
\delta\psi^{(1)}_0=
-\frac{i}{2\Omega^2}\chi^{(1)}
+\half\partial_{\check r}\log \Omega\gamma_{(i)}^r\chi^{(1)}
+\frac{if_0}{2\Omega^6}\chi^{(1)}
-\frac{\Omega^2}{96}G_{\check p\check q\check r}
\gamma_{(i)}^{pqr}\chi^{(2)}=0,\\
\delta\psi^{(2)}_0=
\frac{i}{2\Omega^2}\chi^{(2)}{}^\odot
+\half\partial_{\check r}\log \Omega\gamma_{(i)}^r\chi^{(2)}{}^\odot
+\frac{if_0}{2\Omega^6}\chi^{(2)}{}^\odot
-\frac{\Omega^2}{96}G_{\check p\check q\check r}\gamma_{(i)}^{pqr}
\chi^{(1)}{}^\odot=0.
\eeqa

Similarly, the gravitino variations with $M=5,\cdots,10$ lead to
\beqa\label{gravvarb}
\delta\psi^{(1)}_{\check m}=
D_{\check m}\chi^{(1)}
+\half\partial_{\check m}\log \Omega\chi^{(1)}
-\half\partial_{\check r}\log \Omega\gamma_{(i)}{}_m{}^r\chi^{(1)}
-&& \nonumber \\
-\frac{if_0}{2\Omega^6}\gamma_{(i)}{}_m\chi^{(1)}
-\frac{\Omega^2}{96}G_{\check p\check q\check r}\gamma_{(i)}{}_m{}^{pqr}\chi^{(2)}
+\frac{3\Omega^2}{32}G_{\check m\check q\check r}\gamma_{(i)}{}^{qr}\chi^{(2)}
&=&0 \\
\delta\psi^{(2)}_{\check m}=
D_{\check m}\chi^{(2)}{}^\odot
+\half\partial_{\check m}\log \Omega\chi^{(2)}{}^\odot
-\half\partial_{\check r}\log \Omega\gamma_{(i)}{}_m{}^r\chi^{(2)}{}^\odot
-&& \nonumber \\
-\frac{if_0}{2\Omega^6}\gamma_{(i)}{}_m\chi^{(2)}{}^\odot 
-\frac{\Omega^2}{96}G_{\check p\check q\check r}\gamma_{(i)}{}_m{}^{pqr}
\chi^{(1)}{}^\odot
+\frac{3\Omega^2}{32}G_{\check m\check q\check r}\gamma_{(i)}{}^{qr}
\chi^{(1)}{}^\odot
&=&0
\eeqa

\section{Known solutions}

\subsection{The $T^{1,1}$ solution}

The $T^{1,1}$ space is the intersection of the conifold 
$z_1^2+z_2^2+z_3^2+z_4^2=0$ with the unit sphere. This can be obtained
by applying $SO(3)\times U(1)$ transformations on vectors of the
form
\beq
(z_1^{(0)},\cdots,z_4^{(0)})=(1,i\cos\theta,0,i\sin\theta).
\eeq
We can use the $SO(3)\times U(1)$ rotations to ensure that
$\cos \theta, \sin \theta \ge 0$, and so the manifold is covered if one takes
 $0 \le \theta \le {\pi \over 2}$.  Applying infinitesimal transformations
\beq
\left(\begin{array}{cccc}
1+i\,d\phi  & -\sigma^3   & \sigma^2    & 0 \\
\sigma_3    & 1+i\,d\phi  & -\sigma^1   & 0 \\
-\sigma^2   & \sigma^1    & 1+i\,d\phi  & 0 \\
0           &             & 0           & 1+i\,d\phi
\end{array}\right)
\eeq
leads to 
\beq
d\vec{z}=\left(\begin{array}{c} dz_1 \\ dz_2 \\ dz_3 \\ dz_4 \end{array}\right)=
\left(\begin{array}{c} 
i\,d\phi-i\cos\theta\,\sigma^3 \\
\sigma^3-\cos\theta\,d\phi-i\sin\theta\,d\theta\\
-\sigma^2+i\cos\theta\,\sigma^1 \\
-\sin\theta\,d\phi+i\cos\theta\,d\theta
\end{array}\right).
\eeq
The metric then takes the form \cite{Candelas:1989js}
\beq
ds^2=|d\vec{z}|^2-\frac{1}{6}|\vec{z}\,{}^*\cdot d\vec{z}|^2.
\eeq
The corresponding vielbein is\footnote{We inserted the $-$ sign in $e^5$ 
for later convenience.}
\beqa
e^1&=&\sqrt{\frac{f_0}{3}}\cos\theta\,\sigma^1, \\
e^2&=&\sqrt{\frac{f_0}{3}}\,\sigma^2, \\
e^3&=&\frac{\sqrt{f_0}}{3}\sqrt{3+\cos^2\theta}
\left(\sigma^3-\frac{4\cos\theta}{3+\cos^2\theta}\,d\phi\right), \\
e^4&=&\sqrt{\frac{f_0}{3}}\,d\theta, \\
e^5&=&-\sqrt{\frac{f_0}{3}}\,\frac{2\sin\theta}{\sqrt{3+\cos^2\theta}}\,d\phi,
\eeqa
with a warp factor 
\beq
\Omega^2=\sqrt{f_0}.
\eeq
All the other fields are of course vanishing.

\subsection{The Pilch-Warner fixed point solution}

The vielbein in the Pilch-Warner solution \cite{Pilch:2000ej} is
\beqa
e^1&=&\sqrt{\frac{f_0}{3}}\cos\theta\,\sigma^1, \\
e^2&=&\sqrt{\frac{f_0}{3}}\cos\theta\,\sigma^2, \\
e^3&=&\frac{\sqrt{2f_0}}{3}\cos\theta\,
\sqrt{\frac{3-\cos^2\theta}{2-\cos^2\theta}}
\left(\sigma^3-\frac{2}{3-\cos^2\theta}\,d\phi\right), \\
e^4&=&\sqrt{\frac{2f_0}{3}}\sqrt{2-\cos^2\theta}\,d\theta, \\
e^5&=&-2\sqrt{\frac{f_0}{3}}\sin\theta\,
\sqrt{\frac{2-\cos^2\theta}{3-\cos^2\theta}}\,d\phi
\eeqa
and the warp factor is
\beq
\Omega^2=\sqrt{f_0}\sqrt{2-\cos^2\theta}.
\eeq
Again one has complete coverage of the $S^5$ by the
action of the $SU(2) \times U(1)$ if one takes $0 \le \theta \le {\pi \over 2}$.
The $\ZZ_2$ that reduces the manifold to $S^5/\ZZ_2$ lives
inside the $SU(2)$ and so does not change the range of 
$\theta$.

This set of frames differs from the one in \cite{Pilch:2000ej}  by a shift 
$\sigma^3\mapsto\sigma^3-2d\phi$. This shift is useful so as to make
the assignment of the four-dimensional R-charge more transparent. In this
frame the three-form flux is invariant under the four-dimensional 
R-symmetry.

\subsection{Realization of the $\ZZ_2$ symmetry}

The theory of a single D3-brane probe is the reduction of the 
$SU(N)\times SU(N)$ gauge theory to a gauge theory with a single diagonal 
$U(1)$. One can see that the $\ZZ_2$ symmetry acts as
\beq
(z_1,z_2,z_3,z_4)\mapsto (-z_1,-z_2,-z_3,z_4)
\eeq
on the geometry. This corresponds to a shift $\varphi_3\mapsto\varphi_3+\pi$ 
in the third Euler angle. This symmetry preserves the
$SO(3)$ orbits. Based upon the field theory analysis, we expect the interpolating 
solutions to have  the same property, {\it i.e.}  the geometric action of the $\ZZ_2$ 
symmetry will be implemented in the same way.


The action of the diffeomorphism on the vielbein is 
\beq
(e^1,e^2,e^3,e^4,e^5)\mapsto (-e^1,-e^2,e^3,e^4,e^5).
\eeq
This can be 
undone by a local Lorentz rotation in the 1-2 plane by $\pi$. 
Since in the field theory the $\ZZ_2$ symmetry is a charge conjugation,
the type IIB realization has to contain $S^2$, which is world sheet 
orientation reversal. However, this acts on the $SL(2,\BR)/U(1)$ coset 
fields as $V^\pm_\alpha\mapsto -V^\pm_\alpha$. This has to be undone by 
a type IIB R-symmetry rotation by $\pi$.

The Pilch-Warner solution respects the same $\ZZ_2$ 
symmetry, which is consistent with the $\ZZ_2$ action in the field theory dual.
This  further supports the expectation that this $\ZZ_2$ will indeed be a symmetry
of the complete interpolating family.

\section{The interpolating solutions}

\subsection{Restrictions of the symmetries on the Ansatz}

The most general five-dimensional metric respecting the $SU(2)\times U(1)_R$
symmetry is an $SU(2)\times U(1)$ fibration over an interval.  Using 
coordinate reparametrization invariance in the fiber directions, this 
can be brought into the form
\beqa
e^1&=&A_1\, (\sigma^1+C_1\,d\phi+C_2\,d\theta), \\
e^2&=&A_2\, (\sigma^2+D_1\,d\phi+D_2\,d\theta), \\
e^3&=&A_3\, (\sigma^3+B_1\,d\phi+B_2\,d\theta), \\
e^4&=&A_4\,d\theta, \\
e^5&=&A_5\, d\phi,
\eeqa
However, under the $\ZZ_2$ symmetry $\sigma^1$ and $\sigma^2$ are
odd, whereas $\sigma^3$, $d\theta$ and $d\phi$ are invariant. This
constrains the Ansatz to
\beqa
e^1&=&A_1\, \sigma^1, \\
e^2&=&A_2\, \sigma^2, \\
e^3&=&A_3\, (\sigma^3+B_1\,d\phi+B_2\,d\theta), \\
e^4&=&A_4\,d\theta, \\
e^5&=&A_5\, d\phi.
\eeqa
In Appendix B we give the components of the spin  connection for 
this metric.

The most general Ansatz for the three-form flux, $G$, that respects all the 
symmetries is:
\beq\begin{array}{rl}
G&=g_1\,(e^{134}-ie^{234})+g_2\,(e^{145}-ie^{245})+g_3\,(e^{135}-ie^{235})+ \\
&+g_4\,(e^{134}+ie^{234})+g_5\,(e^{145}+ie^{245})+g_6\,(e^{135}+ie^{235})
\end{array}\eeq
and the most general dilaton-axion background respecting all the 
symmetries is
\beq
P=p\,e^4\qquad{\rm and}\qquad Q=0.
\eeq
Note, that the $U(1)$ connection $Q$ has been gauged away.

The $\ZZ_2$ symmetry acts through the diffeomorphism, the local Lorentz 
rotation by $\pi$ and a ten-dimensional R-symmetry rotation by $\pi$ on
the Killing spinor. This imposes a projector on the Killing spinor
\beq
i\gamma^{12}\chi^{(1)}=\chi^{(1)}\qquad{\rm and}\qquad
i\gamma^{12}\chi^{(2)}{}^\odot=\chi^{(2)}{}^\odot.
\eeq
This projection restricts the spinors $\chi^{(1)}$ and $\chi^{(2)}{}^\odot$
to live in the same two-dimensional subspace of the four-dimensional spinor
space.

\subsection{Solving the supersymmetry variations}

\subsubsection{The ``Magical Combination''}

The magical combination 
$2\delta\psi_0^{(\eta)}+\gamma^1\delta\psi_1^{(\eta)}+ 
\gamma^2\delta\psi_2^{(\eta)}$ of the gravitino variation equations 
\cite{Pilch:2004yg}  is independent of all the fluxes, and depends only upon
the metric. This leads to the projector equations:
\beqa
\left(\frac{(A_1A_2)^\prime}{A_4}\gamma^4-iA_3\gamma^3\right)\chi^{(1)}
&=&\frac{2iA_1A_2}{\Omega^2}\chi^{(1)},\\
\left(\frac{(A_1A_2)^\prime}{A_4}\gamma^4-iA_3\gamma^3\right)\chi^{(2)}{}^\odot
&=&-\frac{2iA_1A_2}{\Omega^2}\chi^{(2)}{}^\odot
\eeqa
In order for the foregoing projector equations to have non-trivial solutions, 
the metric coefficients must satisfy the condition:
\beq
A_3^2=\left(\frac{(A_1A_2)^\prime}{A_4}\right)^2+
\left(\frac{2A_1A_2}{\Omega^2}\right)^2.
\eeq
This condition is equivalent to setting:
\beq\label{magical}
\frac{(A_1A_2)^\prime}{A_3A_4}=\cos\alpha\qquad{\rm and}\qquad
\frac{2A_1A_2}{\Omega^2A_3}=\sin\alpha, 
\eeq
for some function, $\alpha(\theta)$.  The Killing spinors then take the form
\beqa\label{susyforma}
\chi^{(1)}&=&\beta_1\, e^{\frac{i}{2}\phi}
\left(\sin\frac{\alpha}{2}\ket{++}-\cos\frac{\alpha}{2}\ket{+-}\right), \\
\label{susyformb}
\chi^{(2)}{}^\odot&=&\beta_2^*\, e^{-\frac{i}{2}\phi}
\left(\sin\frac{\alpha}{2}\ket{++}+\cos\frac{\alpha}{2}\ket{+-}\right),
\eeqa
where the $\pm$'s refer to the helicities of $i \gamma^{12}$ and 
$i \gamma^{34}$ on $X_4$.  
For consistency of the projector equation, the metric coefficients and 
the function $\alpha$ have to satisfy the differential equation
\beq
\half(\Omega^2A_3\sin\alpha)^\prime=A_3A_4\cos\alpha.
\eeq

We will assume in the following that for the interpolating solutions 
both spinors $\chi^{(1)}$ and $\chi^{(2)}{}^\odot$ are non-vanishing
and for this reason are linearly independent.

\subsubsection{The normalization conditions}

After exploiting the second projector equation, we use
normalization conditions for the Killing spinors coming from the 
symmetry algebra of the problem.
The coefficients $\beta_1$ and $\beta_2$ have to satisfy the normalization 
condition (\ref{normc1}):
\beq
|\beta_1|^2+|\beta_2|^2=1.
\eeq
The other nornalization conditions (\ref{normc2}) lead to the equations 
\beq\label{normcond}
\frac{3A_5}{2\Omega^2}=-\cos\alpha\qquad{\rm and}\qquad
\frac{3A_3B_1}{2\Omega^2}=-\sin\alpha\,(|\beta_1|^2-|\beta_2|^2).
\eeq
For a range of $0\le\alpha\le\frac{\pi}{2}$ the vielbein coefficient 
$A_5$ has to be negative\footnote{Note that this is just a convention and 
$\alpha$ can also be chosen in the range $\frac{\pi}{2}\le\alpha\le\pi$.}.

\subsubsection{The dilatino variation}

The vanishing of the dilatino variations implies
\beqa\label{dv1}
g_4&=&\frac{p}{\Omega^2}
\left(\frac{\beta_2}{\beta_1}-\frac{\beta_1^*}{\beta_2^*}\right), \\ 
\label{dv2}
g_5&=&\frac{p}{\Omega^2\tan\alpha}
\left(\frac{\beta_2}{\beta_1}+\frac{\beta_1^*}{\beta_2^*}\right), \\
\label{dv3}
ig_6&=&\frac{p}{\Omega^2\sin\alpha}
\left(\frac{\beta_2}{\beta_1}+\frac{\beta_1^*}{\beta_2^*}\right).
\eeqa
Note that all three expressions have the same phase. This observation 
is important for the reality conditions.

\subsubsection{Reality conditions}

The next big simplification of the problem comes from realizing that 
the fermion variation equations imply strong reality constraints. This
is due to the reality of all the coefficients in the vielbein.
The external gravitino variation equations imply 
\beq
\Im\left(\frac{\beta_2}{\beta_1}g_1\right)=
\Im\left(\frac{\beta_2}{\beta_1}g_2\right)=
\Re\left(\frac{\beta_2}{\beta_1}g_3\right)=0,
\eeq
and the ``anti-magical'' combination $\gamma^1\delta\psi_1^{(\eta)}-
\gamma^2\delta\psi_2^{(\eta)}$ of gravitino variations implies
\beq
B_2=0.
\eeq
The gravitino variation equation $\delta\psi_4^{(1)}$ then turns into 
two differential equations for $\beta_1$, which take the form
\beq
a_{s,1}\frac{\beta_1^\prime}{\beta_1}+a_{s,2}+
a_{s,3}\frac{\beta_2}{\beta_1}=0\,, \qquad s=1,2\,,
\eeq
with real coefficients $a_{s,t}$. One can take a linear combination of those
two equations such that the term proportional to $\frac{\beta_2}{\beta_1}$
vanishes. This implies that the phases of $\beta_1$ and $\beta_2$
do not depend on $\theta$.

One can use the ten-dimensional $U(1)$ R-symmetry to give the same phase
to $\beta_1$ and $\beta_2$. In addition one can multiply the spinors
$\chi^{(\eta)}$ by an arbitrary constant phase. This allows one to take 
$\beta_1$ and $\beta_2$ to be real, and they can be written as:
\beq
\beta_1=\cos\frac{\beta}{2}\qquad{\rm and}\qquad\beta_2=\sin\frac{\beta}{2}.
\eeq
With this form of $\beta_1, \beta_2$, the spinor Ansatz in
(\ref{susyforma}) and (\ref{susyformb}) 
is equivalent to introducing a dielectric projector as in \cite{Pilch:2004yg}.

It also follows that $g_1$ and $g_2$ are real, $g_3$ is imaginary and
from the anti-magical combination it follows that $p$ is real.  This means that 
all the complex functions in the problem become real functions.

In order to proceed with the gravitino variation equations it is useful 
to define the matrices
\beq\begin{array}{l}
L=\cot\alpha\,\gamma^4-i\csc\alpha\,\gamma^3, \\
M=\frac{1}{\sin\beta\,\sin\alpha}\,\gamma^1(\Bid+
\cos\beta\,\sin\alpha\,\gamma^3+i\cos\alpha\,\gamma^{34}), \\
N=\left(\frac{\cos\beta\,\beta^\prime}{2\sin\beta}+
\frac{\cot\alpha\,\alpha^\prime}{2}\right)\Bid+
\frac{\beta^\prime}{2\sin\beta\,\sin\alpha}\gamma^3+
\frac{i\cot\alpha\,\beta^\prime}{2\sin\beta}\gamma^4+
\frac{i\alpha^\prime}{2\sin\alpha}\gamma^{34}.
\end{array}\eeq
These matrices satisfy the identities:
\beqa
i\chi^{(1)}=L\chi^{(1)} & \qquad{\rm and}\qquad & 
-i\chi^{(2)}{}^\odot=L\chi^{(2)}{}^\odot, \\
\chi^{(2)}=M\chi^{(1)} & \qquad{\rm and}\qquad &
\chi^{(1)}{}^\odot=M\chi^{(2)}{}^\odot, \\
\chi^{(1)}{}^\prime=N\chi^{(1)} & \qquad{\rm and}\qquad &
\chi^{(2)}{}^\odot{}^\prime=N\chi^{(2)}{}^\odot.
\eeqa
This enables one to rewrite the gravitino variation equations in the form:
\beq
R\chi^{(1)}=0\qquad{\rm and}\qquad R\chi^{(2)}{}^\odot=0,
\eeq
for some matrix, $R$.   This implies that $R=0$ modulo $i\gamma^{12}=\Bid$, and
so one can read off the gravitino variation equations as the coefficients of 
$\Bid$,  $\gamma^3$, $\gamma^4$, $\gamma^{34}$.

\subsubsection{The external gravitino variation}

The external gravitino variation equations can be solved for 
$\Omega$, $g_1$, $g_2$ and $g_3$
\beqa
\frac{f_0}{\Omega^4}&=&\cos\beta, \\ \label{gv02}
g_1&=&\frac{4\cot\beta\,\Omega^\prime}{\Omega^3A_4}=
\frac{\beta^\prime}{\Omega^2A_4}, \\ \label{gv03}
g_2&=&-\frac{4\sin\beta}{\Omega^4}-
\frac{4\cot\alpha\,\Omega^\prime}{\Omega^3A_4\sin\beta}=
-\frac{4\sin\beta}{\Omega^4}-
\frac{\cot\alpha\,\beta^\prime}{\Omega^2A_4\cos\beta}, \\ \label{gv04}
g_3&=&-\frac{4i\Omega^\prime}{\sin\beta\,\sin\alpha\,\Omega^3A_4}=
-\frac{i\beta^\prime}{\cos\beta\,\sin\alpha\,\Omega^2A_4}.
\eeqa
We will use these expressions for the three-form flux and $\Omega^4$
to simplify the remaining gravitino variation equations.

\subsubsection{The ``anti-magical combination''}

The anti-magical combination $\gamma^1\delta\psi_1^{(\eta)}-
\gamma^2\delta\psi_2^{(\eta)}$ leads to the following equations
\beqa
\frac{A_1}{A_2A_3}-\frac{A_2}{A_1A_3}&=&
\frac{4p\cos\alpha}{\sin^2\beta\,\sin^2\alpha}, \\
\frac{A_1^\prime}{A_1A_4}-\frac{A_2^\prime}{A_2A_4}&=&
2p\left(1-\frac{2}{\sin^2\beta\,\sin^2\alpha}\right), \\ \label{gvam3}
\frac{A_1B_1}{A_2A_5}-\frac{A_2B_1}{A_1A_5}&=&
\frac{4p\cos\beta}{\sin^2\beta\,\sin\alpha}.
\eeqa
Using the normalization conditions, (\ref{normcond}),  one can see that
the last equation is actually redundant.
Those are the only gravitino variations that contain $p$, but no 
$g_1$, $g_2$ or $g_3$. All the other gravitino variations do not 
contain $p$. A vanishing $p$ would imply that $A_1=A_2$, which inevitably
leads to the Pilch-Warner solution.

\subsubsection{The gravitino variations in the third direction}

\beqa\label{gv31}
\frac{A_1}{A_2A_3}+\frac{A_2}{A_1A_3}-\frac{A_3}{A_1A_2}&=&
-\frac{2}{\Omega^2\sin\alpha}-\frac{2\cos\alpha\,\beta^\prime}
{A_4\sin\beta\,\cos\beta\,\sin^2\alpha}, \\
\frac{A_3B_1^\prime}{2A_4A_5}&=&\frac{2\cos\beta}{\Omega^2}+
\frac{\cot\alpha\,\beta^\prime}{A_4\sin\beta},\\
\frac{A_3^\prime}{A_3A_4}&=&-\frac{\cot\alpha}{\Omega^2}-
\frac{(2\cos^2\alpha+\sin^2\beta\,\sin^2\alpha)\beta^\prime}
{2A_4\sin\beta\,\cos\beta\,\sin^2\alpha},
\eeqa

\subsubsection{The gravitino variations in the fourth direction}

\beqa
\frac{\alpha^\prime}{A_4}&=&\frac{3}{\Omega^2}+
\frac{\cot\alpha\,\beta^\prime}{A_4\sin\beta\,\cos\beta},\\
\frac{A_3B_1^\prime}{2A_4A_5}&=&
\frac{2\cos\beta}{\Omega^2}+
\frac{\cos\alpha\,\beta^\prime}{A_4\sin\beta\,\sin\alpha}.
\eeqa

\subsubsection{The gravitino variations in the fifth direction}

\beqa\label{gv51}
\frac{A_1B_1}{A_2A_5}+\frac{A_2B_1}{A_1A_5}&=&
-\frac{2\beta^\prime}{A_4\sin\beta\,\sin\alpha}, \\
\frac{A_5^\prime}{A_5A_4}&=&\frac{2}{A_5\sin\alpha}+\frac{3\cot\alpha}{\Omega^2}-
\frac{(2-\sin^2\beta)\beta^\prime}{2A_4\sin\beta\,\cos\beta},\\
\frac{2\cos\alpha}{A_5}&=&-\frac{3}{\Omega^2}, \\ 
\frac{A_3B_1^\prime}{2A_4A_5}&=&\frac{2\cos\beta}{\Omega^2}+
\frac{\cot\alpha\,\beta^\prime}{A_4\sin\beta}.
\eeqa
The third equation is equivalent to one of the normalization conditions. 
This confirms that the normalization conditions (\ref{normcond}) are 
chosen with the correct normalization constant.

\subsection{The BPS equations}

One can eliminate most variables from the BPS equations and
the normalization conditions (\ref{normcond}). This leaves three 
independent equations for $\alpha$, $\beta$, $\frac{A_1}{A_2}$ and
$\frac{A_4}{\Omega^2}$. For notational simplicity we define
\beq
g=\frac{A_4}{\Omega^2}\qquad{\rm and}\qquad h=\frac{A_1}{A_2}.
\eeq
With these definitions, the BPS equations are:
\beqa\label{bps1}
\left(\log\left(\frac
{g\sin\beta\,\sin^3\alpha\,(h+h^{-1})}
{\cos\alpha\,\beta^\prime}\right)\right)^\prime&=&
2g\cot\alpha, \\ \label{bps2}
\left(\log\frac{h-h^{-1}}{\cos\beta}\right)^\prime&=&
\frac{2\beta^\prime}{\sin\beta\,\cos\beta\,\sin^2\alpha}, \\ \label{bps3}
\left(\log\frac{\cot\beta}{\cos\alpha}\right)^\prime&=&
3g\tan\alpha.
\eeqa
It is straightforward to verify that BPS equations imply the supersymmetries,
the  Bianchi identities and the equations of motion.  Once one has a solution
to this system one can obtain every other field from $g,h$ and $\beta$.  In
Appendix C we have summarized all the equations needed to achieve this.

One can write (\ref{bps1})--(\ref{bps3}) as a strictly first-order system by 
solving (\ref{bps2}) for  $\beta'$ and substituting the results into (\ref{bps1}) to obtain:
\beq\label{bps4}
\left(\log \left( \frac {g\, (h+h^{-1})}{(3 \,g -1)} \,\frac{\sin^2\alpha}{\cos\beta}
\right) \right)'   ~=~  2g\cot\alpha\,.
\eeq
It is also convenient to use this to substitute for $\beta'$ on the 
right-hand side of (\ref{bps2}) to arrive at:
\beq\label{bps5}
\left(\log\frac{h-h^{-1}}{\sin^2 \alpha \,\cos\beta}\right)^\prime  ~=~ 
 -{6\,g  \over \sin \alpha \, \cos \alpha} .
\eeq
We may then take the BPS system to be  (\ref{bps3})--(\ref{bps5}),
and from this we see that there is now at least one obvious 
integral of motion that can be obtained by taking a simple 
linear combination of (\ref{bps3})--(\ref{bps5}) so as to get zero
on the right-hand side.  Indeed, 
\beq\label{iom}
{\cal I}_0 ~\equiv~ - {g^3 \over (3\, g -1)^3}  \, (h - h^{-1})\, (h + h^{-1})^3\,
{\sin^4 \alpha \over \sin^2 \beta \, \cos^2 \beta} 
\eeq
must be constant as a consequence of the BPS equations.

It is unclear whether this system of equations has a simple,
closed form  for its solution.  The results from gauged supergravity
\cite{Corrado:2002wx} suggest that there should be an explicit solution,
but it has so far  eluded us.   In the next section we will discuss the 
two known (KW and PW) solutions  and use numerical methods to show that
the BPS equations lead to a family that interpolates between these
two solutions. 

\section{Solving the BPS equations}

We will not be able to find the general solution to the BPS equations. 
However, we establish the existence of a one parameter family of solutions 
in several different ways. For this purpose it is useful to first 
understand the boundary conditions. This will allow us to count the integration
constants of the BPS equations.  We will find the linear 
perturbation around the $T^{1,1}$ and Pilch-Warner fixed point solutions. 
Furthermore we find the solutions numerically. 

\subsection{Boundary conditions}

The interpolating solutions are given by $\BR \BP^3\times S^1$ fibrations 
over an interval.  Since the family of solutions should involve trading flux 
for the K\"ahler modulus of the blow-up, the generic member of the family should 
have the same topology as $T^{1,1}$.  The size of the two $S^2$'s will 
change as the three-form flux is changed, but the topology will only 
degenerate to the orbifold when one reaches the PW solution.  
This means that the generic member of the family of solutions
should have exactly the same boundary conditions
on the interval as the $T^{1,1}$ metric.  That is, $A_1$ should vanish
at one end of the $\theta$-interval and $A_5$ should vanish at the other end.
This will then properly fix the topology of the $\BR \BP^3\times S^1$ fibration.
Note that the PW solution also satisfies these boundary conditions, 
and furthermore $A_1,A_2$ and $A_3$ all vanish at $\theta = {\pi \over 2}$.
The vanishing of these extra metric functions merely reflects the collapsed
two-cycle in the orbifold.

We also have not yet fixed the reparametrization invariance ($\theta \to
\tilde \theta(\theta)$).  We do this by requiring that  $\alpha$, defined in
(\ref{normcond}), be the  independent variable and we will adopt this choice 
henceforth.  As we will show below, one has $\alpha\in[0,\frac{\pi}{2}]$.   

Consider the end of the interval where $A_5$  vanishes and where,
generically,  the  coefficients $A_1,\cdots,A_4$ and $\Omega^2$ 
are finite. The coefficient $B_1$ is also generically non-vanishing
as the Klebanov-Witten limit suggests. Then equation 
(\ref{normcond}) implies that $\alpha\rightarrow\frac{\pi}{2}$. Assuming that 
$g$ is generic, equation (\ref{bps3}) implies that $\beta\rightarrow 0$ 
and equation (\ref{bps5}) implies that $h\rightarrow 1$.
Assuming that
\beq
\beta\sim\left(\alpha-\frac{\pi}{2}\right)^s\qquad{\rm and}\qquad 
h-h^{-1}\sim\left(\alpha-\frac{\pi}{2}\right)^t\qquad{\rm with}\qquad s,t>0
\eeq
equation (\ref{bps3}) implies $s=3g-1$ and equation (\ref{bps2}) implies
$t=2s$. Equation (\ref{bps1}) is then trivially satisfied in this limit.

The solution is regular if there is no conical singularity and that the fluxes 
behave in a regular way. The vanishing circle at this end of the interval is 
given by the vector field
\beq
\partial_\phi-B_1\sigma_3.
\eeq
There is no deficit angle if the metric coefficients satisfy
\beq
\lim\limits_{\alpha\rightarrow\frac{\pi}{2}}\frac{B_1A_4}{A_5^\prime}\in\ZZ.
\eeq
The two known solutions impliy that $\frac{B_1A_4}{A_5^\prime}\rightarrow -1$,
and other values of this would correspond to different families of solutions.
One can readily check that 
\beq
\frac{B_1A_4}{A_5^\prime}\rightarrow -s.
\eeq
and so we must have $s=1$, $t=2$ and $g\rightarrow\frac{2}{3}$.
Note that the vanishing circle is not an isometry of the geometry. 
The fluxes can behave like scalars, vectors or two-forms in the 4-5 plane.
Regularity of the fluxes requires
\beq
p\rightarrow 0,\quad g_2\rightarrow g_5\rightarrow 0,\quad 
g_1-ig_3\rightarrow 0\quad{\rm and}\quad g_4+ig_6\rightarrow 0.
\eeq
It is easy to see that all of those regularity 
conditions follow from the behaviour of $\alpha$, $\beta$, $g$ and $h$
\beq\label{ahalfpi}
\alpha\rightarrow \frac{\pi}{2},\qquad
\beta\sim c_1\left(\alpha-\frac{\pi}{2}\right),\qquad 
g\sim\frac{2}{3}\qquad{\rm and}\qquad 
h\sim 1+c_2\left(\alpha-\frac{\pi}{2}\right)^2.
\eeq

Since $\beta=0$ at this end of the interval, the Killing spinors
are of ``Becker type''  and so supersymmetric D3-brane probes
should feel no force and this locus  should be a moduli space for such
probes.

At the other end of the interval $A_1$ must vanish and the 
coefficients $A_2,\cdots,A_5$ and $\Omega^2$ are generically non-vanishing, 
and so one must have $h\rightarrow 0$.  Equation (\ref{magical}) implies that 
$\alpha\rightarrow 0$. Equation (\ref{bps3}) implies 
$\beta^\prime\rightarrow 0$. Assuming that $\beta$ and $g$ stay at generic 
finite values, equation (\ref{bps3}) implies
\beq
\beta^\prime\sim\alpha\qquad{\rm and}\qquad h\sim\alpha^s\qquad{\rm with}\qquad
s>0.
\eeq
Equations (\ref{bps1}) and (\ref{bps2}) then imply $s=1$ and 
$g\rightarrow\half$.

The vanishing cycle at this end of the interval is generated by $\sigma_1$.
Absence of a conical singularity requires
\beq
\frac{A_1^\prime}{A_4}\rightarrow\pm 1,
\eeq
The $T^{1,1}$ solution actually has  $\frac{A_1^\prime}{A_4}\rightarrow -1$. The 
condition for the flux to be regular is
\beq
p\rightarrow 0,\quad g_1\rightarrow g_4\rightarrow 0\quad{\rm and}\quad 
A_2(g_2-g_5)\rightarrow iA_3(g_3+g_6).
\eeq
It is easy to see that all of those regularity
conditions follow from the behaviour of $\alpha$, $\beta^\prime$, $g$ and $h$.
\beq\label{azero}
\alpha\rightarrow 0,\qquad
\beta^\prime ~\sim c_3~ \, \alpha,\qquad 
g~\sim~ \frac{1}{2}\qquad{\rm and}\qquad 
h~\sim~  c_4\, \alpha,
\eeq
where
\beq
c_3~=~ -\lim\limits_{\alpha\rightarrow 0}\frac{\sin\beta\,\cos\beta}{2}.
\eeq

At this end of the interval $\beta$ is generically non-zero 
and supersymmetric D3-brane probes should have a non-trivial 
potential.  However,  if they puff up into D5-branes by the 
dielectric effect, such branes might settle into a supersymmetric
configuration  in this  part of the geometry.

It is at this end of the interval that the $\BR P^3$ degenerates 
into an $S^2$ of finite size unless $c_4=h^\prime=\infty$, which happens 
in the PW limit.

\subsection{Integration constants}

Using $\alpha$ as the independent variable, we see that 
(\ref{bps2})--(\ref{bps4}) is a first order system for
three functions, $g, h$ and $\beta$. There are thus, naively,
three constants of integration, which may be thought of as the 
initial values of these functions at one end of the interval.
However, we saw in the last subsection that regularity of the solution
imposes some constraints on these initial conditions: 
We derived the behaviour of $\beta$, $g$ and $h$ on both ends of the 
interval in such a way that the the solution is regular, has the desired 
toplogy, and the BPS equations are satisfied to leading order. 
On each side of the interval this left two integation constants $c_1$, $c_2$
at $\alpha ={\pi \over 2}$ and $c_3$, $c_4$ at $\alpha =0$.   The complete
solution space of the set of BPS equations is thus three dimensional and
regularity at each end of the interval selects a two-dimensional subspace 
at each end.  Two two-dimensional subspaces in three dimensions 
generically intersect in a one-dimensional subspace, and so there 
will be a (real) one-dimensional family of solutions that are regular
at both ends of the interval.

One can refine this argument using the integral of motion, (\ref{iom}).
As we will show below, ${\cal I}_0$ is given by a simple combination
of $c_1$ and $c_2$, and by a simple combination $c_3$ and $c_4$.
Choosing a value of ${\cal I}_0$ reduces the general solution space
to a two-dimensional space and the regular solutions starting at
each end of the interval to two one-dimensional subspaces. These
subspaces generically intersect at a point, and so given a value
of ${\cal I}_0$ one should expect a single solution that is regular at 
both ends of the $\alpha$-interval.  Thus one expects the family of
solutions we seek to be swept out by varying ${\cal I}_0$.  As we
will show below, the explicit numerical solutions precisely bear out
this picture.

One should recall that we did, in fact, expect a complex one-dimensional 
family of solutions.  The reduction to a real one-dimensional space
came about via some of the gauge choices and rotations we made
earlier.   The real one-dimensional solution space can be complexified 
by reintroducing a constant phase $e^{i\varphi}$ 
to the three-form flux and a phase $e^{2i\varphi}$ to the dilaton $P$.
The other two complex moduli of the solution are the integration
constant $\tau_0$ for the gradient equation for the dilaton-axion and the 
two-form flux through the $S^2$ at $\alpha=\pi$.

\subsection{The Klebanov-Witten limit}

For the $T^{1,1}$ solution, one can use equation (\ref{magical}) to 
determine the angle $\alpha$ in terms of $\theta$ and then eliminate
$\theta$. This leads to 
\beq\label{KWfunctions}
\beta=0,\qquad g=\frac{2}{3+\cos^2\alpha}\qquad{\rm and}\qquad
h=\frac{\sqrt{3}\sin\alpha}{\sqrt{3+\cos^2\alpha}}.
\eeq
This limit looks somewhat singular because $\beta=0$. However,
the ratio $\frac{\beta^\prime}{\sin\beta}$ is not singular. It can 
be calculated using equation (\ref{bps2})
\beq
\frac{\beta^\prime}{\sin{\beta}}=
-\frac{3-\cos^2\alpha}{3+\cos^2\alpha}\,\tan\alpha\,.
\eeq
It is then easy to see that the equations (\ref{bps1}) and (\ref{bps3})
are satisfied.   The integral of motion, (\ref{iom}), diverges and 
corresponds to the singular limit,  ${\cal I}_0 = \infty$.

In order to see that the Klebanov-Witten limit is a smooth limit, one
can do some linearized analysis. Because $\beta$ is vanishing, 
one can expand the BPS equation in
$\delta\left(\frac{\beta^\prime}{\sin\beta}\right)$,
$\delta g$ and $\delta h$. In these variables the
linearized BPS equations turn into a second order system together
with a first order equation
\beq
\delta\beta^\prime=
-\frac{3-\cos^2\alpha}{3+\cos^2\alpha}\,\tan\alpha\,\delta\beta.
\eeq

The obvious solution to the second order system is the trivial one.
The linearized perturbation is then given by 
\beq
\delta\beta=\frac{\cos\alpha}{3+\cos^2\alpha}\,\delta c.
\eeq
where $\delta c$ is a (small) constant of integration.
This solution satisfies all the boundary conditions, especially
$\delta\beta^\prime(\alpha=0)=0$ and 
$\delta\beta\left(\alpha=\frac{\pi}{2}\right)=0$.

It is easy to derive the perturbation of the fields from this. To linear order,
the metric  and the warp factor remain unchanged and the dilaton-axion 
is still zero, however the three-form flux is 
given by:
\beqa
\delta g_1=-\frac{\sin\alpha\,(3-\cos^2\alpha)}{2f_0\,(3+\cos^2\alpha)}
\,\delta c, &&
\delta g_4=\frac{\cos^2\alpha\,\sin\alpha}{f_0\,(3+\cos^2\alpha)}
\,\delta c, \\
\delta g_2=-\frac{\cos\alpha\,(5+\cos^2\alpha)}{2f_0\,(3+\cos^2\alpha)}
\,\delta c, &&
\delta g_5=-\frac{\cos^3\alpha}{f_0\,(3+\cos^2\alpha)}
\,\delta c, \\
\delta g_3=\frac{i(3-\cos^2\alpha)}{2f_0\,(3+\cos^2\alpha)}
\,\delta c, &&
\delta g_6=\frac{i\cos^2\alpha}{f_0\,(3+\cos^2\alpha)}
\,\delta c.
\eeqa
The non-vanishing  $\delta g_4$, $\delta g_5$ and $\delta g_6$ imply 
that at the quadratic order the dilaton-axion becomes non-trivial. It is easy to check that 
this perturbarion satisfies all the boundary conditions.

Since the $T^{1,1}$ solution has no three-form flux and has a trivial 
dilaton-axion background, it is invariant under the phase rotation
$e^{i\varphi}$. For this reason the perturbation can be complexified by 
complexifying $\delta c$.

\subsection{The Pilch-Warner limit}

At the Pilch-Warner fixed point one can show:
\beq\label{PWfunctions}
\cos\beta=\frac{3-\cos^2\alpha}{3+\cos^2\alpha},\quad
\sin\beta=\frac{2\sqrt{3}\cos\alpha}{3+\cos^2\alpha},\quad
g=\frac{2}{3-\cos^2\alpha}\quad{\rm and}\quad
h=1.
\eeq
Again, this limit looks somewhat singular, but equation (\ref{bps2}) defines
the derivative of the logarithm of a vanishing quantity. 
\beq
\left(\log\left(h-h^{-1}\right)\right)^\prime=
-\frac{18+14\cos^4\alpha}
{\sin\alpha\,\cos\alpha\,(3-\cos^2\alpha)(3+\cos^2\alpha)}
\eeq
It is easy to check that the other two BPS equations are satisfied.  
The integral of motion, (\ref{iom}), has the value, ${\cal I}_0 =0$.

As for the Klebanov-Witten limit, one can do a linearized analysis around
the Pilch-Warner point. The BPS equations can be expanded in terms of
$\delta\beta$, $\delta g$ and $\delta\left(\log h\right)^\prime$. Again this 
leads to a second order system together with a first order equation
\beq
\delta h^\prime=
-\frac{18+14\cos^4\alpha}
{\sin\alpha\,\cos\alpha\,(3-\cos^2\alpha)(3+\cos^2\alpha)}
\delta h.
\eeq
Again, the third order system can be solved by the trivial solution. The 
linearized perturbation is then given by
\beq
\delta h=
\frac{\cos^2\alpha\,(3-\cos^2\alpha)^2}{\sin^4\alpha\,(3+\cos^2\alpha)}
\,\delta c, 
\eeq
where $\delta c$ is a (small) integration constant.
This perturbation vanishes at $\alpha=\frac{\pi}{2}$ and diverges at 
$\alpha=0$. The divergence is due to the fact that this
perturbation generates a resolution of the $A_1$ singularity
in the Pilch-Warner geometry.  A  very similar behavior occurs if one
perturbatively expands the resolution  of the $A_1$ singularity 
in the Eguchi-Hanson geometry. This is discussed in Appendix \ref{appa1}. 

The non-vanishing  perturbations of the vielbein coefficients are given by
\beqa
\delta A_1&=&A_1\frac{\delta h}{\sqrt{2}}, \\
\delta A_2&=&-A_1\frac{\delta h}{\sqrt{2}}.
\eeqa
This solution has a similar behavior as the blowup of an $A_1$ singularity,
however, it is geometrically not the same because there are non-zero fluxes and 
curvatures. The sign of $\delta A_1$ suggests that this perturbation
makes $A_1$ vanish at $\theta=\frac{\pi}{2}$ whereas $A_2$ and $A_3$ stay 
finite. Also, the perturbations $\delta p$, $\delta g_4$, $\delta g_5$ and
$\delta g_6$ are non-vanishing, which shows that the interpolating solutions 
indeed have a non-trivial dilaton-axion.

Since the Pilch-Warner fixed point solution has a non-trivial three-form flux,
it is not invariant under the phase rotation $e^{i\varphi}$. For this reason
the foregoing perturbation can be complexified by
\beq
\delta g_i=ig_i\delta\varphi.
\eeq

\subsection{The round $S^5/\ZZ_2$}\label{spheresolution}

Another very simple solution to the BPS equations is given by:
\beq
\beta=0,\qquad g=1\qquad{\rm and}\qquad h=1.
\eeq
It is easy to check that this is actually the round $S^5/\ZZ_2$.
The regularity of the metric at $\alpha=\frac{\pi}{2}$ implies that
$\phi$ has a periodicity of $3\pi$. For this reason volume integrals have 
an extra factor of $\frac{3}{2}$. This is important for the central 
charge calculations in the next section.

\subsection{Numerical solutions}

We now set about obtaining numerical solutions to the system of equations
(\ref{bps1})--(\ref{bps3}), and we will indeed see that this system of equations leads 
to a family of solutions  that interpolates between the Pilch-Warner and $T^{1,1}$
geometries.   As in the previous section, we fix the freedom to 
reparametrize the $\theta$-coordinate, $\theta \to \tilde \theta (\theta)$, 
by  taking $\alpha \in [0,{\pi \over 2}]$ to be the independent variable.
The next step is to use (\ref{bps2})--(\ref{bps4}) to obtain expressions for $g'$, $h'$ and 
$\beta'$  in terms $g,h$ and $\beta$.  One can then employ a simple Euler method to 
get the numerical solution once one has specified ``initial velocities'' for $g,h$ and $\beta$.  
{\it A priori} there are three constants of integration, but as we 
described earlier, regularity reduces this to a one parameter family of solutions
parametrized by the value of ${\cal I}_0$. 

We find the solutions for the functions $g,h$ and $\beta$ by ``shooting,'' that is, we 
vary initial data at $\alpha = 0$
and adjust it so as to hit the proper values at $\alpha = {\pi \over 2}$.  In particular,
we make use of the asymptotics  given in (\ref{azero}) and (\ref{ahalfpi}).  
At $\alpha =0$ one has $\beta' = 0$ and $h= 0$ and so the equations in (\ref{bps1})
appear to be somewhat singular, however a careful series expansion about
$\alpha = 0$ leads to a regular expansion of all the undetermined functions,
and one finds: 
\beqa
h &=& c_4 \,  \alpha ~+~ O(\alpha^3)  \,, \qquad g~=~ \half ~+~ 
\frac{1}{16} (4\,c_4^2 -1) \, \alpha^2 ~+~ O(\alpha^4) \,, \\ 
 \beta &=& \beta_0 ~-~  \frac{1}{4}\, \sin \beta_0  \, \cos \beta_0\,  
\alpha^2 ~+~ O(\alpha^4)   \,.
\eeqa
Similarly, at $\alpha = {\pi \over 2}$ one finds:
\beqa
h &=& 1~+~ c_2 \,  \Big(\alpha - {\pi \over 2}\Big)^2 ~+~ 
O\Big(\Big(\alpha - {\pi \over 2}\Big)^4\Big)  \,, \\
\qquad g &=&{2 \over 3}  ~+~ \frac{1}{9} (3\,c_1^2 -2) \,
\Big(\alpha - {\pi \over 2}\Big)^2 ~+~  O\Big(\Big(\alpha - {\pi \over 2}\Big)^4\Big) \,, \\ 
 \beta &=&c_1 \, \Big(\alpha - {\pi \over 2}\Big) ~-~  \frac{1}{6}\, c_1\,   (3 - c_1^2) \, 
 \Big(\alpha - {\pi \over 2}\Big)^3 ~+~ O\Big(\Big(\alpha - {\pi \over 2}\Big)^5\Big)   \,.
\eeqa
There are thus two free parameters at either end of the interval:
 $\beta_0$ and $c_4$ at $\alpha = 0$ and  $c_1$ and $c_2$ at $\alpha = 
 {\pi \over 2}$.  One can use the series expansions to check that the 
 constant of the motion, (\ref{iom}), is given by
 \beq\label{numiom}
 {\cal I}_0 ~=~ {1 \over c_4^4 \,  \sin^2 \beta_0  \, \cos^2 \beta_0} ~=~
 -\frac{128}{27} \, {c_2 \over c_1^2}.
 \eeq

It is simplest to shoot from $\alpha = 0$ where the value of
$\beta_0$ is chosen so as to select the particular member of the family of 
solutions  and then the value of $c_4$  is adjusted so that one arrives at 
$g=\frac{2}{3}$ as $\alpha\rightarrow\frac{\pi}{2}$.   We use the series expansion
at  $\alpha = 0$  (evolved to fairly high order) to start the numerical solution, 
and then simply use an Euler method to generate the complete solution.  
By choosing $c_4$ to arrive at $g=\frac{2}{3}$ one finds that the asymptotic 
behavior of all the three functions obeys the proper asymptotics at
$\alpha = {\pi \over 2}$.  

The functions $g,h$ and $\beta$ for the KW and PW solutions are given by
(\ref{KWfunctions})  and  (\ref{PWfunctions}).   In particular, observe that for
$\alpha = 0$ one has $\beta_0 = {\pi \over 3}$ for the PW solution.  
We therefore found the numerical solutions for  several values of
$\beta_0 $ in the range $0 \le \beta_0 \le {\pi \over 3}$.  The results for
$\beta$, $g$ and $h$ are plotted in Figures 1, 2 and 3.   We have also 
plotted the exact results for the KW and PW solutions.  
The value of the integral of motion, (\ref{numiom}), monotonically
increases across the family from $0$ for the PW solution to  infinity
for the KW solution.  It is clear from these
graphs that the solutions to the BPS equations do indeed interpolate 
between the KW and PW solutions, and that there is a smooth family of
solutions in which the flux of the PW solution is traded for a blowing-up
of the non-trivial two-cycle.  

\begin{figure}[bth]
\centerline{ \epsfig{file=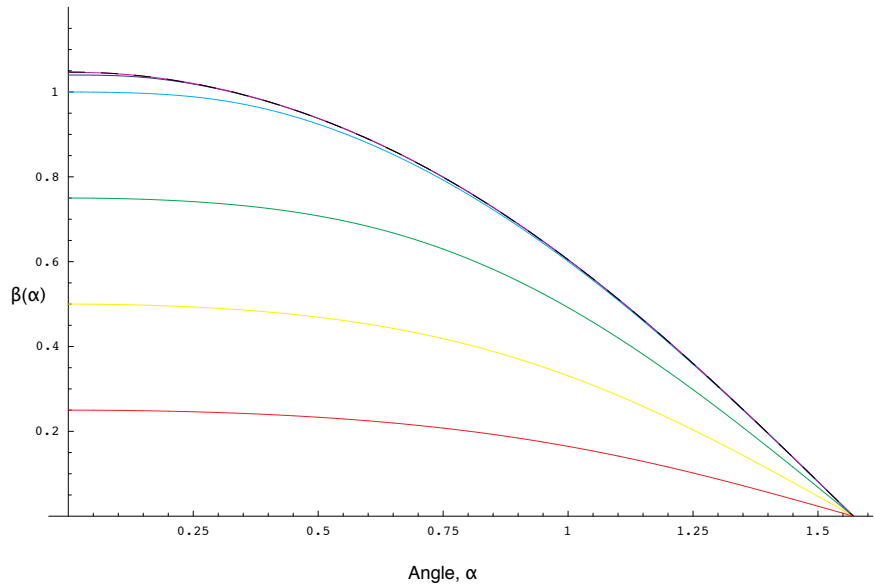,width=12cm}}
\caption{\sl Plots of the function $\beta(\alpha)$ against  $\alpha$
for of six different values of the initial data. The curves merge into 
a dashed enveloping curve that shows $\beta(\alpha)$ for the 
Pilch-Warner solution.  For the Klebanov-Witten solution one has $\beta \equiv 0$. }
\label{betagraphs}
\end{figure}
\begin{figure}[bth]
\centerline{ \epsfig{file=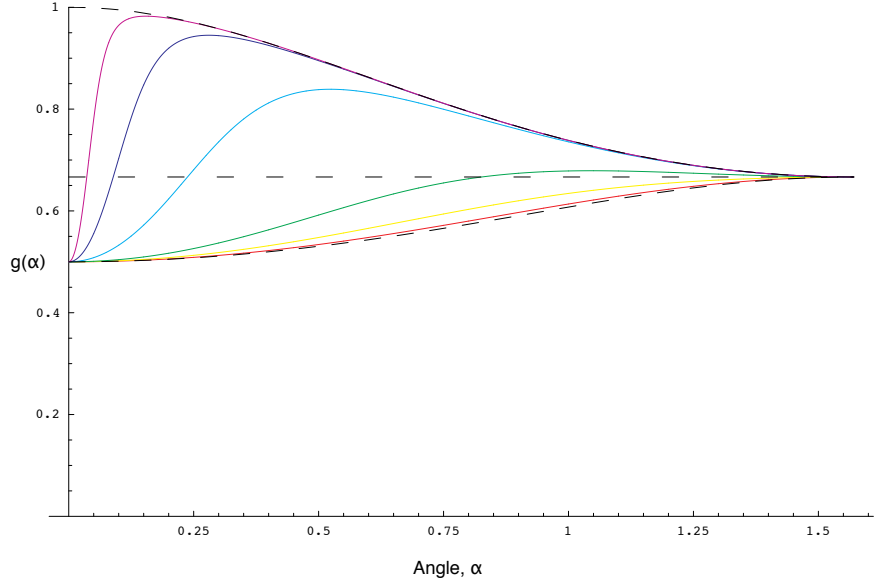,width=12cm}}
\caption{\sl Plots of the function $g(\alpha)$ against  $\alpha$
for of six different values of the initial data.  The upper
and lower dashed lines show the function, $g(\alpha)$, for
the Pilch-Warner solution and for the Klebanov-Witten solution respectively.
The horizontal dashed line shows the ``target value'', ${2 \over 3}$,
for $g(\alpha)$ at $\alpha = \frac{\pi}{2}$. }
\label{ggraphs}
\end{figure}
\begin{figure}[bth]
\centerline{ \epsfig{file=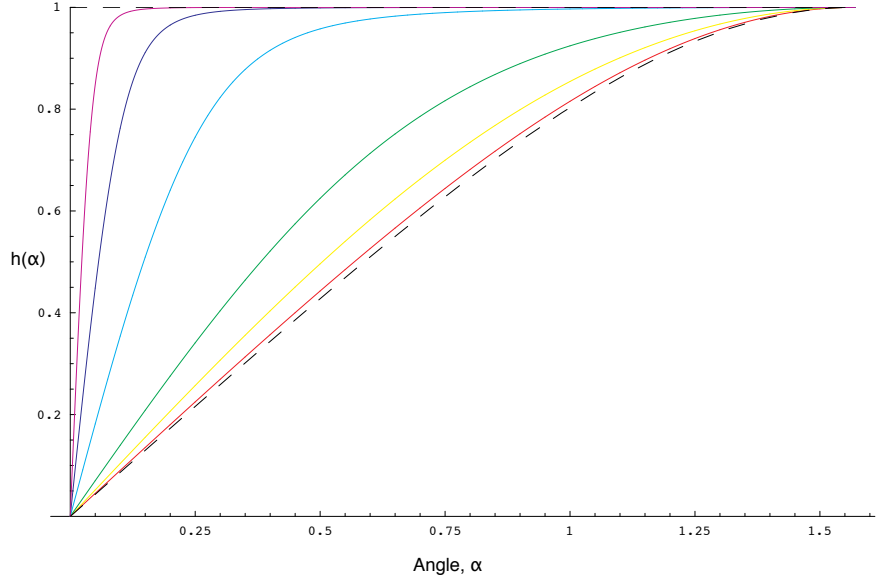,width=12cm}}
\caption{\sl Plots of the function $h(\alpha)$ against  $\alpha$
for of six different values of the initial data.  The lower dashed
curve shows $h(\alpha)$ for the Klebanov-Witten solution, while 
$h(\alpha) \equiv 1$ for the Pilch-Warner solution.}
\label{hgraphs}
\end{figure}

We have focussed on the set of regular solutions to the BPS equations.
As we noted earlier, there is a three-parameter family of solutions in general.
Our numerical solutions show that the other solutions to the BPS equations
can be characterized as solutions starting from $g={1 \over 2}$ at
$\alpha =0$ and arriving at some arbitrary value of $g$ at  $\alpha ={\pi \over 2}$.
The absence of conical singularities required that $(3g -1) \in \ZZ$, and 
the Klebanov-Witten solution imposed $(3g -1) = 1$.  However, there might
be other interesting solutions to our BPS equations that are regular geometries
but with different asymptotic values of $g$.

\section{The central charge}

As a final check on our results, we calculate the central charge of the family
of solutions. It turns out that this is actually an exact calculation
even though the exact solutions are not known. The central charge
of the holographic dual gauge theory is proportional to the effective
five-dimensional Newton constant \cite{Henningson:1998gx,Gubser:1998vd}.

\subsection{Calculating the effective five-dimensional Newton constant}

The effective five-dimensional Newton constant $G_5$ is given by
\beq
G_5=G_{10}\int\limits_{X_5}\frac{1}{\Omega^2}\,
e^1\wedge e^2\wedge e^3\wedge e^4\wedge e^5
\eeq
Using the vielbein Ansatz this can be reduced to
\beq
G_5=2(2\pi)^3G_{10}\int\limits_I\frac{A_1A_2A_3A_4A_5}{\Omega^2}\,d\theta
\eeq
Using the equations (\ref{magical}), (\ref{normcond}) one can show that
\beq
G_5=-\frac{2(2\pi)^3G_{10}}{3}\int\limits_I(A_1^2A_2^2)^\prime\,d\theta.
\eeq

For the family of $\CN=1$ theories the boundary conditions imply that
\beq\label{fptc}
G_5^{(IR)}=\frac{2(2\pi)^3G_{10}f_0^2}{27},
\eeq
whereas for the $\CN=2$ theory 
\beq
G_5^{(UV)}=\frac{3(2\pi)^3G_{10}f_0^2}{48}.
\eeq
Note that the factor 3 in the enumerator is due to the different periodicity
of $\phi$ as discussed in section \ref{spheresolution}.
These formulas depend on the integration constant, $f_0$. This constant
was introduced as the coefficient of the five-form flux. For this reason 
it is related to the number $N_{D3}$ of D3-branes, which is the rank 
of the gauge group.  However, (\ref{fptc}) already shows that the central
charge of the dual field theory is constant across the entire family of solutions,
and independent of the choice of the initial data for the BPS equations.
To conclude that this implies that the central charge of the dual field 
theory is constant across the family one  really needs to show that 
the parameter, $f_0$, represents the number of $D3$-branes present
in the family of solutions.  While this seems highly plausible,
we  will now prove it.

\subsection{Calculating the rank of the gauge group}

The Bianchi identity \cite{Schwarz:1983qr}
\beq
dF^{(5)}=\frac{i}{8}G\wedge G^\ast
\eeq
implies that the five-form flux is not only sourced by D3-branes, but also
by three-form flux. For this reason the total five-form flux cannot be used to 
determine the rank of the gauge group. The effect of the 
three-form flux can be subtracted as follows
\beq
N_{D3}=\int\limits_{X_5}
\left(F^{(5)}-\frac{i}{16}\epsilon_{\alpha\beta}A^\alpha\wedge F^\beta\right).
\eeq
The five-form under the integral is not gauge invariant by itself, but
the integral is gauge invariant.

To determine this integral, we need to relate the quantities appearing here 
to metric and field coefficients. The internal part of the field strength $F^{(5)}$ is 
given by
\beq
F^{(5)}_{int}=\frac{f_0A_1A_2A_3A_4A_5}{\Omega^{10}}\,
\sigma^1\wedge\sigma^2\wedge\sigma^3\wedge d\theta\wedge d\phi
\eeq

The three-form flux $F^{(3)}=F^1=(F^2)^\ast$ is related to $G$ by
\beq
G=-\epsilon_{\alpha\beta}V^\alpha_+\,F^\beta.
\eeq
Using the identity
\beq
|V_+^2|^2-|V_+^1|^2=1,
\eeq
the foregoing relation can be inverted to yield
\beq
F^{(3)}=V_+^2{}^\ast G+V^1_+G^\ast.
\eeq
In our geometry $G$ has the form
\beq\begin{array}{rl}
G=&h_1\sigma^1\wedge\sigma^3\wedge d\theta+
h_2\sigma^2\wedge\sigma^3\wedge d\theta+
h_3\sigma^1\wedge\sigma^3\wedge d\phi+\\
+&h_4\sigma^2\wedge\sigma^3\wedge d\phi+
h_5\sigma^1\wedge d\theta\wedge d\phi+
h_6\sigma^2\wedge d\theta\wedge d\phi,
\end{array}\eeq
which implies, that $F^{(3)}$ has the form
\beq\begin{array}{rl}
F^{(3)}&=f_1\sigma^1\wedge\sigma^3\wedge d\theta+
f_2\sigma^2\wedge\sigma^3\wedge d\theta+
f_3\sigma^1\wedge\sigma^3\wedge d\phi+\\
+&f_4\sigma^2\wedge\sigma^3\wedge d\phi+
f_5\sigma^1\wedge d\theta\wedge d\phi+
f_6\sigma^2\wedge d\theta\wedge d\phi.
\end{array}\eeq
The field strength, $F^{(3)}$, satisfies the Bianchi identity $dF^{(3)}=0$, which implies
\beq
f_5=-f_4^\prime\qquad{\rm and}\qquad f_6=f_3^\prime.
\eeq
A two-form potential $A^{(2)}$ for such an $F^{(3)}$ is then
\beq
A^{(2)}=-f_1\sigma^2\wedge d\theta+f_2\sigma^1\wedge d\theta-
f_3\sigma^2\wedge d\phi+f_4\sigma^1\wedge d\phi.
\eeq

This can be used to determine
\beq
\epsilon_{\alpha\beta}A^\alpha F^\beta=
2(f_1f_3^\ast-f_1^\ast f_3+f_2f_4^\ast-f_2^\ast f_4)\,
\sigma^1\wedge\sigma^2\wedge\sigma^3\wedge d\theta\wedge d\phi
\eeq
or
\beq
\epsilon_{\alpha\beta}A^\alpha F^\beta=
2(h_1h_3^\ast-h_1^\ast h_3+h_2h_4^\ast-h_2^\ast h_4)\,
\sigma^1\wedge\sigma^2\wedge\sigma^3\wedge d\theta\wedge d\phi
\eeq
This can be reexpressed in terms of the vielbein coefficients
\beq\begin{array}{c}
\epsilon_{\alpha\beta}A^\alpha F^\beta=\\
=-4A_3^2A_4A_5((g_1+g_4)(g_3+g_6)A_1^2+(g_1-g_4)(g_3-g_6)A_2^2)\,
\sigma^1\wedge\sigma^2\wedge\sigma^3\wedge d\theta\wedge d\phi
\end{array}\eeq

One can check that
\beq\begin{array}{l}
-\frac{(A_1^2A_2^2\cos^2\beta)^\prime}{3f_0}= \\
\frac{f_0A_1A_2A_3A_4A_5}{\Omega^{10}}+
\frac{iA_3^2A_4A_5((g_1+g_4)(g_3+g_6)A_1^2+(g_1-g_4)(g_3-g_6)A_2^2)}{144}.
\end{array}\eeq
which implies that
\beq
N_{D3}=-\frac{2(2\pi)^3}{3f_0}\int\limits_I(A_1^2A_2^2\cos^2\beta)^\prime\,d\theta.
\eeq

For the family of $\CN=1$ theories this yields:
\beq
N_{D3}^{(IR)}=\frac{2(2\pi)^3f_0}{27}
\eeq
and for the $\CN=2$ theory this is
\beq
N_{D3}^{(UV)}=\frac{3(2\pi)^3f_0}{48}.
\eeq
This enables us to express the effective five-dimensional Newton constant in terms 
of the rank of the gauge group
\beq
G_5^{(IR)}=\frac{27G_{10}N_{D3}^{(IR)}{}^2}{2(2\pi)^3}\qquad{\rm and}\qquad 
G_5^{(UV)}=\frac{48G_{10}N_{D3}^{(IR)}{}^2}{3(2\pi)^3}.
\eeq
The ratio of the effective five-dimensional Newton constants is exactly the
ratio of the central charges of the UV and the IR gauge theories
\beq
\frac{G_5^{(IR)}}{G_5^{(UV)}}=\frac{27}{32}=\frac{c_{(IR)}}{c_{(UV)}}.
\eeq
Thus the family of solutions has precisely the correct central charge
to be the duals of the family of fixed points predicted in 
\cite{Corrado:2002wx,Corrado:2004bz}.

\section{Conclusions}

We have found the long-sought family of $AdS_5$ vacuum solutions that
interpolate between the $T^{1,1}$ compactification and the flux
compactification of Pilch and Warner \cite{Pilch:2000ej}.   This family
of solutions is holographically dual to the family of $\CN=1^*$ 
IR fixed points that can be obtained by flowing from an $\CN=2$,
$\ZZ_2$ quiver gauge theory.  In the field theory, this family is
parametrized by the ratio, $m_1/m_2$, of masses given to the
chiral multiplets on each node of the quiver.  In supergravity the
difference of the masses, $m_1- m_2$, is dual to the K\"ahler modulus
of a non-trivial $S^2$, while the sum of the masses, $m_1+ m_2$, is dual 
to a non-trivial, three-form field strength.   Thus the family represents a kind of
continuous geometric transition in which a K\"ahler deformation is
traded for flux.

One of the surprises, and perhaps one of the reasons why this solution
was not discovered earlier, is that the generic solution has a non-trivial
dilaton.  It is surprising because the dilaton background is trivial for the
two previously know (KW and PW) solutions.   There are obvious questions
about whether there is any interesting physics to be learned from the
non-trivial dilaton profiles.  On the more mathematical side, it raises questions
about the underlying geometric structure of these solutions.  One of the
important insights of \cite{Halmagyi:2004jy} was that the geometry of the PW solution,
and indeed the flows to and around it \cite{Freedman:1999gp,Gowdigere:2005wq}, 
possessed an integrable complex structure, and indeed were ``almost Calabi-Yau.''
The non-trivial dilaton profile, and indeed the fact that it is real, seems 
to be at odds with the integrability of the complex structure.  We have tried
the obvious generalizations of the integrable complex structure found in 
\cite{Halmagyi:2004jy} and they fail to work here, and this failure perhaps explains
the incompatibility of the complex structures, noted in \cite{Gowdigere:2005wq}, of 
the PW  flow  and of the Calabi-Yau metric that must underlie \cite{Halmagyi:2004jy}  
the KW flow.  There is thus an interesting issue as to how to characterize
the geometry of the interpolating family obtained here.

The system of BPS equations that we obtained were surprisingly complicated,
also probably as a consequence of the non-trivial dilaton profile.  
This is all the more surprising in the light of the results of \cite{Corrado:2002wx}
that led to the conjectured existence of the family of solutions.  It was
shown in \cite{Corrado:2002wx} that, from the perspective of five-dimensional,
$\CN=4$ gauged  supergravity, all the vacuum solutions in the family, and indeed 
all the flows to them,  were governed by exactly the same set of equations.
The complete family, in five-dimensional supergravity, is swept out by the action of 
an $SU(2)$ symmetry.  One would therefore, naively, expect an equally simple
formulation in ten-dimensions.  However, as was pointed out in \cite{Corrado:2002wx},
and as we see explicitly here, this sweeping out of the family involves some
extremely non-trivial trading of very different geometric quantities in
ten dimensions.  It is certainly not the first time that a trivial symmetry in
lower dimensions has led to subtle or profound effects in higher 
dimensions, and indeed the parallels between the present example
and mirror symmetry are rather intriguing.   It would certainly be very
interesting to find how the symmetry that sweeps out the family acts in
ten dimensions.   This might be similar to the $SL(2,\BR)$ 
action in \cite{Lunin:2005jy}. For this reason there should be a simpler 
form of our BPS equations and a way to solve them analytically. 
However, in string theory such a continous symmetry group of
the supergravity will be broken down to a discrete duality group 
by solitonic excitations \cite{Hull:1994ys,Witten:1995ex,Halmagyi:2004ju}. 

There is also the issue of the flow solutions: We have found the fixed points,
but it would be very useful to find the family of flows from the quiver gauge
theories to these fixed points.  Finding these might also shed light upon
the underlying geometric structure.

As a final comment, we found the family of solutions by a very careful analysis of 
the symmetries of the field theory.  In particular, the discrete $\ZZ_2$ symmetry
in combination with the $SO(3)$ symmetry 
played a very significant role in fixing the metric Ansatz and in determining one of
the supersymmetry projectors.  We suspect that such a careful treatment of such
discrete symmetries of will also give new insights into how to solve other open 
problems in holographic descriptions of field theories, especially
for field theories related to $\CN=4$ SYM.  

\begin{acknowledgments} 
\nopagebreak

\noindent This work is supported in part by funds provided by the DOE under grant  
number DE-FG03-84ER-40168. The work of NH is supported in part by a Fletcher  
Jones Graduate Fellowship from USC. 
Research at the Perimeter Institute is supported in part by funds from NSERC of Canada.

NH would like to thank the theory group at Stony Brook and ANU for hospitality. 
CR and NW would like to thank the 
Aspen Center for Physics in which part of the work was done.

We would like to thank Andy Brandhuber, Alex Buchel, Rich Corrado, 
Jerome Gauntlett, Jaume Gomis, Peter Mayr, Andrei Starinets and Nemani
Suryanarayana for useful discussions.
\end{acknowledgments}

\begin{appendix}

\section{Some Clifford algebra}

\subsection{Generalities}

The Clifford algebra is defined by the anticommutation relations
\beq
\{\gamma^m,\gamma^n\}=2\eta^{mn},
\eeq
where $\eta^{mn}=\eta^m\delta^{mn}$. We choose a representation in which 
$\sqrt{\eta^m}\gamma^m$ is Hermitean\footnote{By the square root we mean 
$\sqrt{1}=1$ and $\sqrt{-1}=i$.}. Given a complex structure, one can 
define the raising and lowering operators
\beq
\Gamma^m=\sqrt{\eta^{2m-1}}\gamma^{2m-1}+i\sqrt{\eta^{2m}}\gamma^{2m},
\quad{\rm and}\quad (\Gamma^m)^\dagger=
\sqrt{\eta^{2m-1}}\gamma^{2m-1}-i\sqrt{\eta^{2m}}\gamma^{2m}.
\eeq
Then the raising and lowering operators satisfy the following anticommutation 
relations:
\beq
\{\Gamma^m,\Gamma^n\}=\{(\Gamma^m)^\dagger,(\Gamma^n)^\dagger\}=0
\quad{\rm and}\quad \{\Gamma^m,(\Gamma^n)^\dagger\}=4\delta^{mn}.
\eeq
One can then define the fermion number operators
\beq
F^m=i\sqrt{\eta^{2m-1}}\gamma^{2m-1}\sqrt{\eta^{2m}}\gamma^{2m}=
1-\half\Gamma^m(\Gamma^m)^\dagger=-1+\half(\Gamma^m)^\dagger\Gamma^m.
\eeq
The chirality operator is then the product of all the Fermion number operators
$\gamma=F^1\cdots F^n$.

The Fermion number operators have eigenvalues $\pm 1$. The eigenvalues of
the Fermion number operators can be used to label a basis of states.
One can define a ground state $\ket{0}$ which is anihilated by all the 
lowering operators. It has Fermion number $-1$ for all Fermion number 
operators. All other states can be gotten by applying raising operators.
If one labels a state by $\ket{\nu_1,\cdots,\nu_n}$, then the raising and
lowering operators act as follows:
\beqa
\ket{\nu_1,\cdots,+1,\cdots,\nu_n}&=&
\half\nu_1\cdots\nu_{m-1}(\Gamma^m)^\dagger
\ket{\nu_1,\cdots,-1,\cdots,\nu_n},\\
\ket{\nu_1,\cdots,-1,\cdots,\nu_n}&=&
\half\nu_1\cdots\nu_{m-1}\Gamma^m
\ket{\nu_1,\cdots,+1,\cdots,\nu_n}.
\eeqa
This defines the matrix elements of the gamma matrices. One can see that in 
this basis $\Gamma^m$ is real. From this follows that
\begin{itemize}
\item The matrices $\sqrt{\eta^m}\gamma^m$ are Hermitean,
\item The matrices $\sqrt{\eta^{2m-1}}\gamma^{2m-1}$ are symmetric and real and
\item The matrices $\sqrt{\eta^{2m}}\gamma^{2m}$ are antisymmetric and 
imaginary.
\end{itemize}

In general there are matrices $B$, $C$ and $D$ such that
\beqa\label{Bdefn}
(\gamma^m)^*&=&\eta_BB\gamma^m B^{-1},\\
(\gamma^m)^\dagger&=&C\gamma^m C^{-1},\\
(\gamma^m)^t&=&\eta_BD\gamma^m D^{-1},
\eeqa
where $\eta_B=\pm 1$ is a constant which is chosen (if possible) such that
$BB^*=1$. One can see that $D=(B^{\dagger})^{-1} C$. Given a spinor $\epsilon$, 
$\epsilon^\odot=B^{-1}\epsilon^*$, $\bar\epsilon=\epsilon^\dagger C$ and 
$\tilde\epsilon=\epsilon^t D$ transform covariantly.

If $BB^*=1$ one can impose the Majorana condition $\epsilon=B^{-1}\epsilon^*$.
And if $B$ commutes with the chirality operator $\gamma$, one can impose 
the Majorana-Weyl condition.

In the following we collect useful Gamma matrix identities in various dimensions.

\subsection{$Spin(1,9)$}

Chirality operator:
\beq
\gamma=-\gamma^{0\cdots 9}
\eeq
Complex conjugation:
\beq
B=\gamma^{013579}
\eeq
\beq
B\gamma^MB^{-1}=-(\gamma^M)^*
\eeq
\beq
B\gamma_{(10)}B^{-1}=\gamma_{(10)}^*
\eeq
\beq
BB^*=1
\eeq
Hermitean conjugation:
\beq
C=\gamma^0
\eeq
\beq
C\gamma^MC^{-1}=(\gamma^M)^\dagger
\eeq
Transpose:
\beq
D=(B^{\dagger})^{-1}\, C=-\gamma^{13579}
\eeq
\beq
D\gamma^MD^{-1}=-(\gamma^M)^t
\eeq

\subsection{$Spin(1,4)$}

Chirality operator:
\beq
\gamma^4=-\gamma^{0123}
\eeq
\beq
\gamma^{01234}=1
\eeq
Complex Conjugation:
\beq
B=\gamma^{013}
\eeq
\beq\label{Bfivedefn}
B\gamma^\mu B^{-1}=(\gamma^\mu)^*
\eeq
\beq
BB^*=-1
\eeq
Hermitean conjugation:
\beq
C=\gamma^0
\eeq
\beq
C\gamma^\mu C^{-1}=(\gamma^\mu)^\dagger
\eeq
Transpose:
\beq
D=(B^{\dagger})^{-1} \, C=\gamma^{13}
\eeq
\beq
D\gamma^\mu D^{-1}=(\gamma^\mu)^t
\eeq

\subsection{$Spin(5)$}

Chirality operator:
\beq
\gamma^5=-\gamma^{1234}
\eeq
\beq
\gamma^{12345}=-1
\eeq
Complex Conjugation:
\beq
B=\gamma^{24}
\eeq
\beq
B\gamma^m B^{-1}=(\gamma^m)^*
\eeq
\beq
BB^*=-1
\eeq
Hermitean conjugation:
\beq
C=1
\eeq
\beq
C\gamma^mC^{-1}=(\gamma^m)^\dagger
\eeq
Transpose:
\beq
D=(B^{\dagger})^{-1} \, C=\gamma^{24}
\eeq
\beq
D\gamma^mD^{-1}=(\gamma^m)^t
\eeq

It is easy to check that
\beq
\begin{array}{cc}
B^{-1}\ket{++}^*=\ket{--}, & B^{-1}\ket{+-}^*=-\ket{-+}, \\
B^{-1}\ket{-+}^*=\ket{+-}, & B^{-1}\ket{--}^*=-\ket{++}.
\end{array}
\eeq

\subsection{Decomposition of a ten-dimensional spinor}

We want to decompose spinors in ten-dimensional Minkowski space of mostly
minus signature into four-dimensional and six-dimensional spinors. The gamma
matrices can be decomposed as
\beq
\gamma_{(10)}^\mu=
\left(\begin{array}{cc}
0 & \gamma_{(e)}^\mu\otimes 1_{(i)} \\
\gamma_{(e)}^\mu\otimes 1_{(i)} & 0
\end{array}\right)
\quad{\rm and}\quad
\gamma_{(10)}^m=
\left(\begin{array}{cc}
0 & -1_{(e)}\otimes\gamma_{(i)}^m \\
1_{(e)}\otimes\gamma_{(i)}^m & 0
\end{array}\right).
\eeq
Note that the internal gamma matrices $\gamma_{(i)}^m$ have a $+$-signaturte. 

The ten-dimensional chirality operator is given by
\beq
\gamma_{(10)}=-\gamma_{(10)}^{0\cdots 9}=
\left(\begin{array}{cc} 1 & 0 \\ 0 & -1 \end{array}\right),
\eeq
the complex conjugation is given by
\beq
B_{(10)}=
\left(\begin{array}{cc}
B_{(e)}\otimes B_{(i)} & 0 \\ 0 & -B_{(e)}\otimes B_{(i)}
\end{array}\right)
\eeq
and the hermitean conjugation is given by
\beq
C_{(10)}=
\left(\begin{array}{cc}
0 & C_{(e)}\otimes C_{(i)} \\ C_{(e)}\otimes C_{(i)} & 0
\end{array}\right).
\eeq

\section{The spin connection of the internal metric}

The derivatives of the vielbein are
\beqa
de^1&=&\frac{A_1^\prime}{A_1A_4}e^4\wedge e^1+\frac{A_1}{A_2A_3}e^2\wedge e^3-
\frac{A_1B_1}{A_2A_5}e^2\wedge e^5-\frac{A_1B_2}{A_2A_4}e^2\wedge e^4, \\
de^2&=&\frac{A_2^\prime}{A_2A_4}e^4\wedge e^2-\frac{A_2}{A_1A_3}e^1\wedge e^3+
\frac{A_2B_1}{A_1A_5}e^1\wedge e^5+\frac{A_2B_2}{A_1A_4}e^1\wedge e^4, \\
de^3&=&\frac{A_3^\prime}{A_3A_4}e^4\wedge e^3+\frac{A_3}{A_1A_2}e^1\wedge e^2+
\frac{A_3B_1^\prime}{A_4A_5}e^4\wedge e^5, \\
de^4&=&0, \\
de^5&=&\frac{A_5^\prime}{A_5A_4}e^4\wedge e^5.
\eeqa
This leads to the following spin connection:
\beqa
\omega_{114}&=&\frac{A_1^\prime}{A_1A_4}, \\
\omega_{123}&=&-\frac{A_1}{2A_2A_3}+\frac{A_2}{2A_1A_3}+\frac{A_3}{2A_1A_2}, \\
\omega_{124}&=&\frac{A_1B_2}{2A_2A_4}-\frac{A_2B_2}{2A_1A_4}, \\
\omega_{125}&=&\frac{A_1B_1}{2A_2A_5}-\frac{A_2B_1}{2A_1A_5}, \\
\omega_{224}&=&\frac{A_2^\prime}{A_2A_4}, \\
\omega_{213}&=&-\frac{A_1}{2A_2A_3}+\frac{A_2}{2A_1A_3}-\frac{A_3}{2A_1A_2}, \\
\omega_{214}&=&\frac{A_1B_2}{2A_2A_4}-\frac{A_2B_2}{2A_1A_4}, \\
\omega_{215}&=&\frac{A_1B_1}{2A_2A_5}-\frac{A_2B_1}{2A_1A_5}, \\
\omega_{312}&=&\frac{A_1}{2A_2A_3}+\frac{A_2}{2A_1A_3}-\frac{A_3}{2A_1A_2}, \\
\omega_{334}&=&\frac{A_3^\prime}{A_3A_4}, \\
\omega_{345}&=&-\frac{A_3B_1^\prime}{2A_4A_5}, \\
\omega_{412}&=&-\frac{A_1B_2}{2A_2A_4}-\frac{A_2B_2}{2A_1A_4}, \\
\omega_{435}&=&-\frac{A_3B_1^\prime}{2A_4A_5}, \\
\omega_{512}&=&-\frac{A_1B_1}{2A_2A_5}-\frac{A_2B_1}{2A_1A_5}, \\
\omega_{534}&=&\frac{A_3B_1^\prime}{2A_4A_5}, \\
\omega_{554}&=&\frac{A_5^\prime}{A_5A_4}.
\eeqa

\section{Recovering the fields}

Going through all the independent BPS equations one can recover the 
vielbein coefficients from $\alpha$, $\beta$, $g$ and $h$
\beqa
\Omega^2&=&\sqrt\frac{f_0}{\cos\beta}, \\
A_1&=&\sqrt{-\frac{f_0\sin^3\alpha\,\sin\beta\,g(h^2+1)}
{4\beta^\prime\cos\alpha}}, \\
A_2&=&\sqrt{-\frac{f_0\sin^3\alpha\,\sin\beta\,g(h^2+1)}
{4\beta^\prime\cos\alpha\,h^2}}, \\
A_3&=&-\frac{\sqrt{f_0\cos\beta}\sin^2\alpha\,\sin\beta\,g(h^2+1)}
{2\beta^\prime\cos\alpha\,h}, \\
A_4&=&g\sqrt{\frac{f_0}{\cos\beta}}, \\
A_5&=&-\frac{2}{3}\sqrt{\frac{f_0}{\cos\beta}}\cos\alpha, \\
B_1&=&\frac{4\beta^\prime\cos\alpha\,h}{3g\sin\alpha\,\sin\beta\,(h^2+1)}, \\
p&=&-\frac{\beta^\prime\sin\beta\,(h^2-1)}{2\sqrt{f_0\cos\beta}\,g(h^2+1)}, \\
g_1&=&\frac{\cos\beta\,\beta^\prime}{f_0g}, \\
g_2&=&-\frac{4\sin\beta\,\cos\beta}{f_0}-
\frac{\cot\alpha\,\beta^\prime}{f_0g}, \\
g_3&=&-\frac{i\beta^\prime}{f_0g\sin\alpha}, \\
g_4&=&\frac{\cos\beta\,\beta^\prime(h^2-1)}{f_0g(h^2+1)}, \\
g_5&=&-\frac{\beta^\prime\cot\alpha\,(h^2-1)}{f_0g(h^2+1)}, \\
g_6&=&\frac{i\beta^\prime(h^2-1)}{f_0g\sin\alpha\,(h^2+1)}.
\eeqa

\section{The resolution of an $A_1$ singularity}\label{appa1}

The Eguchi-Hansen metric can be written as
\beq
ds^2=4r^6(a+r^4)^{-\frac{5}{2}}(2a+r^4)dr^2+
(a+r^4)^{-\half}\left(\frac{r^8}{2a+r^4}(\sigma^1)^2+
(2a+r^4)((\sigma^2)^2+(\sigma^3)^2)\right),
\eeq
with $r\ge 0$. This is a global coordinate system which allows a
smooth $a\rightarrow 0$ limit. A corresponding vielbein is
\beqa
e^1&=&A_1\sigma^1=r^4(a+r^4)^{-\frac{1}{4}}(2a+r^4)^{-\half}\sigma^1,\\
e^2&=&A_2\sigma^2=(a+r^4)^{-\frac{1}{4}}(2a+r^4)^\half\sigma^2,\\
e^3&=&A_3\sigma^3=(a+r^4)^{-\frac{1}{4}}(2a+r^4)^\half\sigma^3,\\
e^4&=&A_4\sigma^4=2r^3(a+r^4)^{-\frac{5}{4}}(2a+r^4)^\half dr.
\eeqa
The linearized perturbation around $a=0$ is given by
\beqa
\delta A_1&=&-\frac{5}{4r^3}\delta a, \\
\delta A_2=\delta A_3&=&\frac{3}{4r^3}\delta a, \\
\delta A_4&=&\frac{1}{2r^4}\delta a
\eeqa
The size of the deformation can be determined using the natural metric
\beq
\delta g_{ij}\delta g_{kl} g^{ik}g^{jl}=
\sqrt{A_1A_2A_3A_4}\,\sum_i\left(\frac{2\delta A_i}{A_i}\right)^2=
\sqrt{2}r^{-\frac{13}{2}}\delta a^2.
\eeq
This diverges as $r\rightarrow 0$, which indicates that the range of 
validity of the linearized approximation is smaller for small $r$.

\end{appendix}

\vfill
\pagebreak
 
\bibliographystyle{utphys}
\bibliography{adsfamily}
\end{document}